\documentclass[11pt,twocolumn]{article}

\usepackage{fullpage}
\usepackage{amsfonts,amssymb,amsmath}
\usepackage{tikz,pgfplots}
\usepackage{graphicx}
\usepackage{float}
\usepackage{subcaption}
\usepackage{hyperref} 
\usepackage{multicol}
\usepackage[utf8]{inputenc}
\usepackage{cite}
\usepackage{authblk}

\usepackage{amsmath,amssymb,physics}
\usepackage{bm}        
\usepackage{newtxtext,newtxmath}

\usepackage{graphicx}
\usepackage{subcaption}   
\usepackage{caption}
\usepackage[english]{babel}
\usepackage[a4paper, left=1.3cm, right=1.3cm, top=1.7cm, bottom=2.0cm]{geometry}

\usepackage{xcolor}
\usepackage{soul}         
\usepackage{relsize,scalerel}  
\usepackage{setspace}
\usepackage{bbold}        
\usepackage{lipsum}

\usepackage{pgf}
\usepackage{pgfpages}

\usepackage{easyReview}

\begin{document}

	\title{High-Fidelity Teleportation of Continuous-Variable Quantum States Via\\ Non-Ideal Qutrit Entangled Resources}

	\author[1]{Fatemeh~Taghipoor}
	\author[1*] {Mojtaba~Golshani  }
	\author[2]{Mostafa~Motamedifar}
	\author[3]{Khatereh~Jafari}
	\affil[1]{Department of interdisciplinary Physics and Technology, Faculty of science, Shahid Bahonar University of Kerman, Kerman, 7616913439, Iran}
	\affil[2]{Department of Physics, Faculty of science, Shahid Bahonar University of Kerman, Kerman, 7616913439, Iran}
	\affil[3]{Center for Quantum Engineering and Photonics Technologies, Sharif University of Technology, Tehran, Iran}
	\affil[*]{Corresponding author: golshani@uk.ac.ir}
	
	\date{}
	

	\maketitle
	
	\twocolumn[
	
	\begin{abstract}
		Achieving near-unity fidelity in conventional continuous-variable quantum teleportation schemes based on two-mode squeezed vacuum states is fundamentally unattainable. To overcome this limitation, alternative approaches utilizing ensembles of two-dimensional entangled qubits have been proposed. In this work, we investigate continuous-variable quantum teleportation employing entangled qutrit resources under realistic noise effects. The results demonstrate that the proposed scheme performs well in both ideal and noisy conditions, enabling high-fidelity teleportation with a reasonable success probability.
		\\
		\\
		\\
		
	\end{abstract}
	]
	

	\section{Introduction}
	
	Quantum technologies, due to their capabilities far exceeding those of classical physics-based technologies, have recently become one of the main subjects of interest for many researchers~\cite{m0,m3,m4,m5}. Quantum teleportation is one of the most distinguished protocols in quantum information, which has a special place in many new quantum technologies ~\cite{0,01,02}. For instance, it plays a fundamental role in the realization of quantum networks~\cite{1}, distributed quantum computation~\cite{2}, and one-way quantum computation~\cite{3}. This remarkable discovery was first introduced by Bennett \textit{et al.}~\cite{4}.\\
	
	Ideally, quantum teleportation involves the complete transfer of an unknown quantum state from a sender (Alice) to a remote receiver (Bob). This process requires two channels, classical and quantum ones. In the illustrated general protocol for quantum teleportation, it is the quantum channel that plays a central role. The quantum channel consists of a two mode maximally entangled state that is already shared between two remote systems called Alice (A) and Bob (B), and this strong quantum long-range correlation between the two modes enables the teleportation of an unknown quantum state. This is possible by combining the unknown input state with the quantum channel on Alice's side (system A) and by subsequently performing a Bell measurement, whose projective effect transfers the quantum information of the unknown state to Bob’s system (B) via the long-range correlations. The measurement outcome is then announced to Bob through an authenticated classical channel, which could be a telegraph wire, and Bob uses this additional classical information to fully reconstruct the input state from his own system B. This is done by performing a unitary transformation conditioned on the specific Bell measurement outcome. It should be noted that during the Bell measurement process Alice's initial state is destroyed, which is a direct consequence of the no-cloning theorem. So the net result of the teleportation is the removal of the unknown state from Alice's side and its appearance on Bob's side some time later, just after the time it takes for a classical message to go from Alice to Bob~\cite{5}.\\
	
	Depending on whether the Hilbert space of the unknown quantum state has finite or infinite dimensions, quantum teleportation is classified into two categories: discrete variables (DV) and continuous variables (CV). Quantum teleportation was first introduced in 1993 for discrete variables~\cite{4} and then extended to continuous variables by Vaidman in 1994~\cite{6}. The best-known example of discrete-variable teleportation concerns the world of qubits, a quantum system with two distinguishable states, such as spin-$\frac{1}{2}$ particles (e.g., electrons) and the two polarization states of a single photon~\cite{4}. While examples of continuous variable teleportation include states that could be, for example, the radiation modes of an electromagnetic field or the vibrating modes of a mechanical oscillator. Due to the infinite dimensions of their Hilbert space, these states can be described by continuous variables such as field quadratures or position and momentum~\cite{5,15,7}.\\
	
	The latter type of quantum teleportation is particularly interesting, mainly due to experimental advantages, including high compatibility with classical communication infrastructures and the ease of implementing most Gaussian operations in this type of protocol, since it can be implemented with high efficiency using relatively simple states, transformations, and detectors ~\cite{5,8,9}.\\
	Ideally, using a quantum channel such as the maximally entangled EPR pair (in the discrete-variable teleportation) or the maximally entangled two-mode squeezed vacuum state (TMSV), with infinite squeezing (in the continuous-variable teleportation), allows for perfect quantum teleportation, such that the output state is exactly equal to the input state. However, in real situations, Alice and Bob access limited resources, and the output state will be similar to the input state by a quantity called Fidelity (F). The value of this quantity is one for two completely identical states and zero for two completely different (orthogonal) states ~\cite{5}.\\
	Nevertheless, despite the apparent success of CV implementations, in all of them the teleportation fidelity is severely limited, because achieving 100\% fidelity in CV teleportation requires infinite squeezed and therefore unphysical resources, so in principle it is not possible to perform ideal CV teleportation using Gaussian squeezed state resources~\cite{8}.\\
	
	In 2013, Ralph \textit{et al.} proposed a new method for continuous variable teleportation that approaches unit fidelity with finite resources~\cite{8}. In this protocol, instead of using a standard TMSV as the quantum channel, a supply of maximally entangled single-photon states is used in a multimode interferometric setup. They investigated the teleportation scheme for various input states, including coherent states, superposition states, and two-mode squeezed vacuum states, and found several situations where near-unity teleportation fidelity can be achieved with finite resources. The teleportation scheme proposed in ~\cite{8} involves $N$ qubit teleporters (with a 2D Hilbert space) in a multimode interferometer. The input state is divided into $N$ modes, each of which is teleported by a qubit teleporter. The teleported outputs are recombined in an \textit{N}-splitter, and if all photons exit through one port, the teleportation is successful ~\cite{8}.\\
	In this approach, if the states to be teleported have low average energy, high fidelity can be achieved with a small number of $N$. However, if the average photon number of the input state increases, to achieve high fidelity near unity, the number of divisions $N$ needs to be very large, which can affect the success probability of the teleportation process~\cite{8}. Therefore, although the proposed scheme allows for achieving fidelity near unity, it requires a large number of two-dimensional entangled sources and also has a low success probability.\\
	In this paper, we present a solution that, in addition to improving the fidelity of CV quantum teleportation, also achieves an acceptable success probability. This is done by increasing the dimensions of the quantum channel used in each of the teleporters.\\
	In 2005, a scheme for teleporting an unknown two-qutrit entangled state through a three-level GHZ quantum channel was proposed, and then extended to the general state of arbitrary Qudit~\cite{10}. In addition, the teleportation of both an unknown single-qutrit state and a two-qutrit state via a three-level GHZ quantum channel was introduced in ~\cite{16}. In ~\cite{11}, the teleportation of unknown qubit and qutrit states using an entangled pair of qutrits was investigated. It was shown that, by properly constructing the measurment bases, both states can be faithfully transmitted from Alice to Bob. 
	High-dimensional entangled systems, such as qutrits, present fundamental and multifaceted advantages over their two-dimensional qubit counterparts. In quantum communication, they enable higher-information-density coding, significantly boosting channel capacity beyond the limits achievable with qubits and allowing for the transmission of more quantum information. Furthermore, high-dimensional systems offer inherently greater resilience to noise and robustness against eavesdropping strategies. This increased tolerance for error elevates the maximum acceptable disturbance in Quantum Key Distribution (QKD) protocols, enabling secure operation at lower signal-to-noise ratios and potentially extending transmission distances. This superior performance is fundamentally rooted in their ability to span a vastly larger Hilbert space, which not only better preserves the higher-order moments of the input quantum state but also provides a richer resource set. This richness is essential for achieving a quantum computational advantage, as a large entangled Hilbert space is a prerequisite for quantum speedup and enables more efficient and flexible quantum algorithms. Consequently, the transition from qubits to qutrits in protocols like quantum teleportation is not merely an incremental improvement but a paradigm shift, unlocking new potentials for secure, high-capacity, and fault-tolerant quantum information processing~\cite{new1,new2}.
	\\
	In Ref.~\cite{8}, a supply of two-dimensional maximally entangled states (Bell states) is used as a two-dimensional quantum channel (2D) to perform the teleportation. In this paper, we extend the scheme to a three-dimensional (3D) quantum channel, i.e., maximally entangled qutrits, in order to increase the fidelity of the teleportation process without significantly reducing the success probability. Furthermore, since perfect entanglement and an ideal quantum channel cannot be realized in real situations, we have also analyzed the influence of noise and channel imperfections on the teleportation process in the proposed scheme.\\
	In fact, since entanglement is inherently a highly sensitive property, it can be easily degraded by any interaction between the system and its environment, which may introduce noise and decoherence, and may even lead to the so-called "sudden death of entanglement". Many efforts have been made to investigate the effect of various types of noise on various quantum teleportation schemes~\cite{12m,12,13,13p}. For example, in ~\cite{12} the effect of noise on qubit teleportation, and in ~\cite{13} the fidelity of qutrit teleportation in the presence of various types of quantum noise have been investigated.\\
	Since it is unavoidable to investigate the effect of noise in the teleportation process, this paper also investigates the effect of three common quantum noises on the teleportation process in the proposed scheme based on multimode interference. The results of the study show that the proposed protocol remain reasonably robust to noise, showing only the moderate reduction in the performance metrices of the teleportation, under non-ideal conditions.\\
	While the theoretical advantages of high-dimensional quantum teleportation, such as increased fidelity and channel capacity, are evident~\cite{new1,p14}, their practical implementation presents significant challenges, primarily rooted in the generation of high-quality entangled qudit states and the necessity of deterministic high-dimensional Bell-state measurement~\cite{p8,p14,p15}. 
	Experimental efforts utilizing path, time-energy, and orbital angular momentum encoding have successfully demonstrated the feasibility of generating entangled qutrits with high fidelities~\cite{p4,p5,p6,p13,p15}. 
	This progress is driven by advanced techniques like integrated photonics, which offers superior phase stability~\cite{p4,p7,p11}, and specialized pump modulation, which boosts generation rates~\cite{p13}. 
	However, the intrinsic difficulty of achieving a complete high-dimensional Bell-state measurement using only linear optics necessitates more complex solutions~\cite{p8,p14}. 
	Recent achievements have begun to circumvent this obstacle by utilizing ancillary entanglement, where auxiliary entangled photon pairs are introduced to facilitate deterministic high-dimensional Bell-state measurement~\cite{p14}, a technique demonstrated for qutrits with clear potential for scalability to even higher dimensions~\cite{p8,p14}. 
	The convergence of these advances is further supported by remarkable progress in integrated silicon photonics, which provides robust, scalable, and highly controllable platforms for both generating and analyzing multidimensional entanglement~\cite{p7,p11}. 
	Critically, these modern platforms are capable of performing the arbitrary unitary operations essential for the final stages of teleportation~\cite{p4,p11}, thereby effectively bridging the gap between theoretical protocol and experimental realization and solidifying the practical relevance of high-dimensional quantum information processing for future technologies~\cite{p7,p11,p14}.
	\\
	The structure of this paper is as follows. In Section~\ref{section2}, the proposed scheme is investigated in the ideal case and in the absence of quantum noise. Section~\ref{section3} includes an investigation of the real case and different quantum noise effects on fidelity and probability of success in the proposed teleportation scheme. Finally, Section~\ref{section4} includes a summary and conclusion.

	\label{section2}
	\section{Continuous-variable teleportation in multimode interference-based scheme: Ideal case}
	
\begin{figure}[h!]
		\centering
		\begin{tabular}[b]{c}
			\includegraphics[width=1\linewidth]{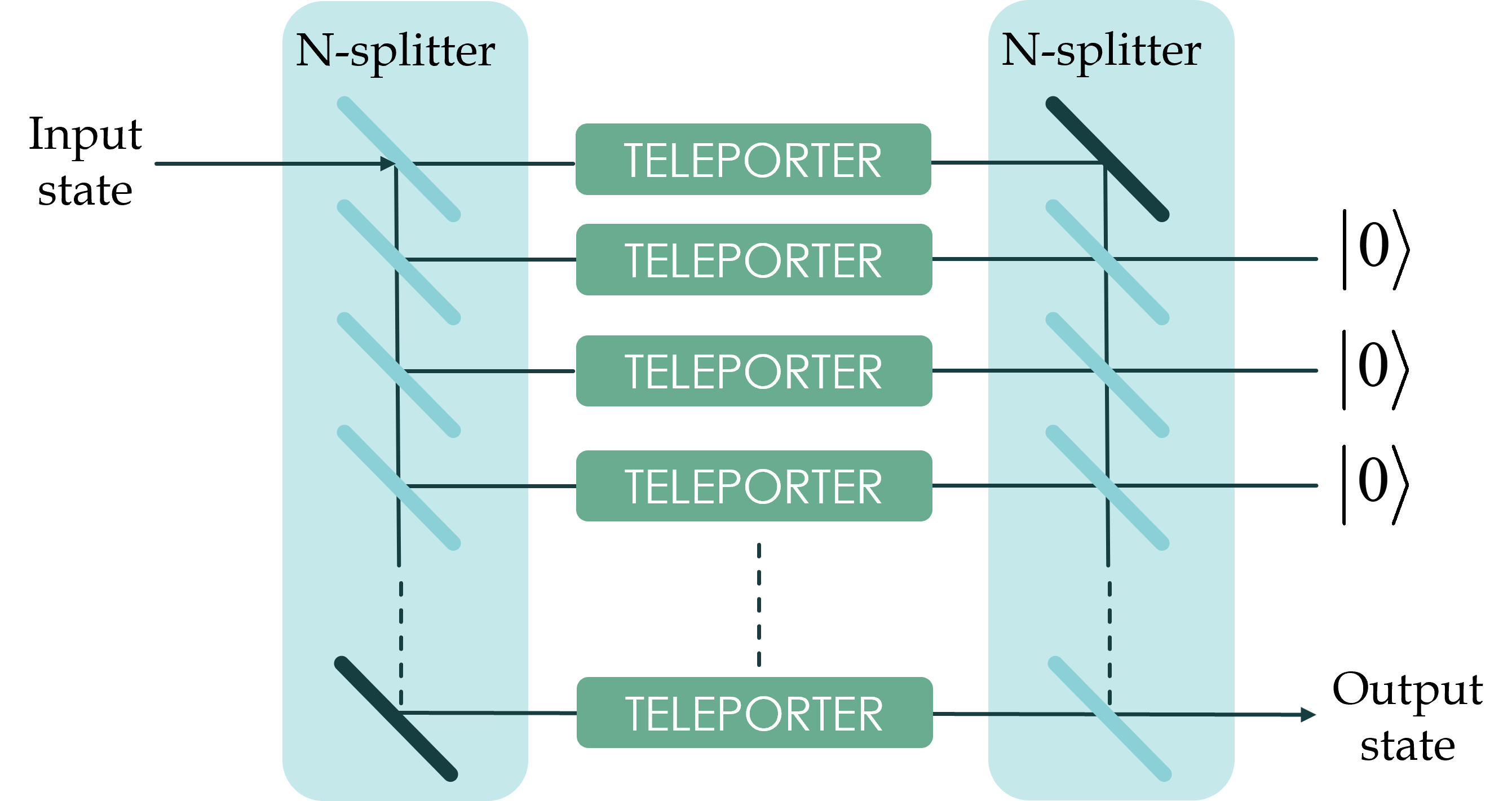}\\
		\end{tabular}
		\caption{\footnotesize{
				Schematic of the teleportation scheme, including $N$ qutrit teleporters in a multimode interferometer, the input state is split into $N$ states, each of which is teleported by a qutrit teleporter. The teleported outputs are recombined in an \textit{N}-splitter, and teleportation is successful if all photons exit through a single port.
		}}
		\label{fig:nsplitter}
	\end{figure}

	The proposed scheme for improving the fidelity of CV teleportation is illustrated in Fig.~\eqref{fig:nsplitter}. This scheme is similar to the one proposed in~\cite{8}. The difference is that the teleporters of each arm of the \textit{N}-splitter perform the qutrit teleportation process as compared to the qubit teleportation. \\
	
	As we will see later, this increase in the dimension of the teleporters from 2D to 3D improves the fidelity, increases the success prabability, and reduces the number of \textit{N}-splitter arms. In this protocol, in the ideal case, the quantum channel of each 3D teleporter is a two-mode entangled qutrit state, as follows :
	\begin{eqnarray}
		\label{1}
		\left| \Phi  \right\rangle  = \frac{1}{{\sqrt 3 }}(\left| {00} \right\rangle  + \left| {11} \right\rangle  + \left| {22} \right\rangle )\,.
	\end{eqnarray}
	Initially, we investigate teleportation of a coherent state $\left| \alpha  \right\rangle  = {e^{ - \frac{{{{\left| \alpha  \right|}^2}}}{2}}}\sum\limits_{k = 0}^\infty  {\frac{{{\alpha ^k}}}{{\sqrt {k!} }}} \left| k \right\rangle $.
	As shown in Fig.~\ref{fig:nsplitter}, this state enters \textit{N}-splitter. Each output of this \textit{N}-splitter is the following attenuated coherent state:
	\begin{eqnarray}
		\label{2}
		\left| {\frac{\alpha }{{\sqrt N }}} \right\rangle  = {e^{\frac{{ - {{\left| \alpha  \right|}^2}}}{{2N}}}}(\left| 0 \right\rangle  + \frac{\alpha }{{\sqrt N }}\left| 1 \right\rangle  + \frac{{{\alpha ^2}}}{{\sqrt {2!} N}}\left| 2 \right\rangle  + ...)\,.
	\end{eqnarray}
	At this step, to carry out the teleportation, these states are combined with the maximally entangled states of Eq.~\eqref{1} to form the state ${\left| {\frac{\alpha }{{\sqrt N }}} \right\rangle ^{ \otimes N}} \otimes {\left| \Phi  \right\rangle ^{ \otimes N}}$ and are then teleported using one of the generalized Bell projectors in 3D space and conditional unitary transformations~\cite{11}. This reduces the Hilbert space of individual states to three dimensions, and so the teleported states can be written as
	\begin{eqnarray}
		\label{3}
		{\left| {\frac{\alpha }{{\sqrt N }}} \right\rangle ^{ \otimes N}} \to {\left[ {{e^{ - \frac{{{\alpha ^2}}}{{2N}}}}(1 + \frac{\alpha }{{\sqrt N }}{{\hat a}^\dag } + \frac{{{\alpha ^2}}}{{2!N}}(\hat{a}^\dagger)^2)\left| 0 \right\rangle } \right]^{ \otimes N}}\,.
	\end{eqnarray}
	After the teleportation, the states are coherently combined in an \textit{N}-splitter by projecting all but one output in the vacuum state. The final output state is
	\begin{eqnarray}
		\label{7}
		\left| {{\psi _{\text{out}}}} \right\rangle  = \frac{1}{{\sqrt {{P_\text{s}}} }}\sum\limits_{m = 0}^{2N} {{C_m}\left| m \right\rangle }\, ,
	\end{eqnarray}
	\begin{eqnarray}
		\label{8}
		\begin{aligned}
			{C_m} &= {e^{ - \frac{{{{\left| \alpha  \right|}^2}}}{2}}}{\left( {\frac{\alpha }{N}} \right)^m}N!\sqrt {m!}\\ 
			&\times\mathop{\mathlarger{\mathlarger{\sum}}}\limits_{k = \max \{ 0,m - N\} }^{\left\lfloor {{\raise0.7ex\hbox{$m$} \!\mathord{\left/
							{\vphantom {m 2}}\right.\kern-\nulldelimiterspace}
						\!\lower0.7ex\hbox{$2$}}} \right\rfloor } {\frac{1}{{k!(m - 2k)!(N - m + k)!}}} \,\, , 
		\end{aligned}
	\end{eqnarray}
	where
	\begin{eqnarray}
		\label{9}
		{P_\text{s}} = \sum\limits_{m = 0}^{2N} {{{\left| {{C_m}} \right|}^2}} \, ,
	\end{eqnarray}
	is the success probability of the teleportation process (the condition where each of the teleporters only teleports the Fock-state components of the input state with up to three photons, and at the output of the final \textit{N}-splitter, all photons exit from a single port). Since this teleportation is not perfect due to the reduction of the Hilbert space dimension, the fidelity between the input coherent state and the output teleported state is given by
	\begin{eqnarray}
		\label{10}
		{F} = {\left| {\left\langle \alpha  \right|\left. {{\psi _{\text{out}}}} \right\rangle } \right|^2} = \frac{{{e^{ - {{\left| \alpha  \right|}^2}}}}}{{{P_\text{s}}}}{\left| {\sum\limits_{m = 0}^{2N} {\frac{{{{({\alpha ^ * })}^m}}}{{\sqrt {m!} }}{C_m}} } \right|^2}.
	\end{eqnarray}\\
	Fidelity measures the similarity between two quantum states on a scale from 0 to 1, with 0 indicating no similarity and 1 indicating that the states are identical.
	\\
	\begin{figure}[h!]
		\centering
		\begin{subfigure}{0.7\linewidth}
			\centering
			\includegraphics[width=\linewidth]{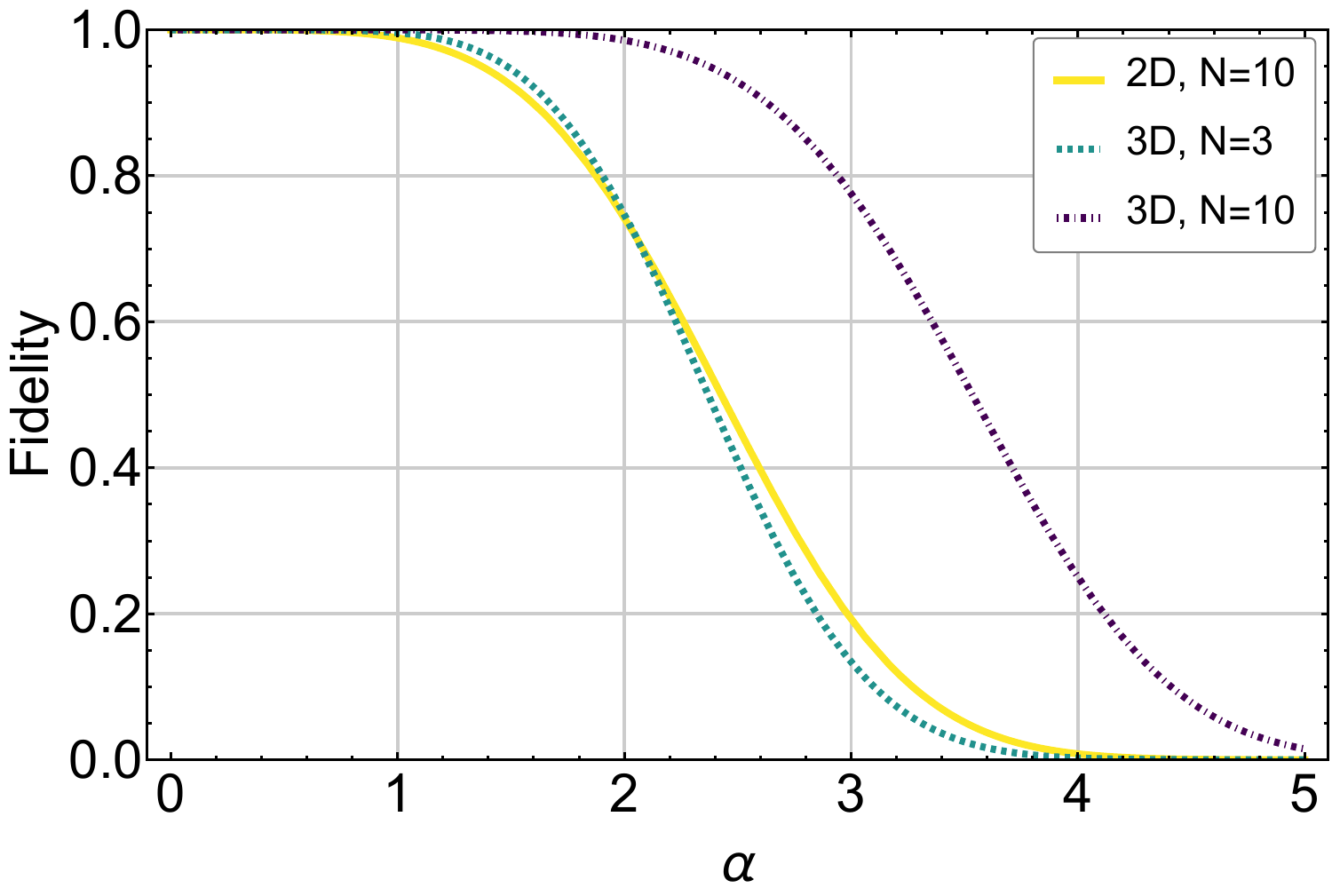}
			\caption{(a)}
			\label{fig2-1}
		\end{subfigure}
		\qquad
		\begin{subfigure}{0.7\linewidth}
			\centering
			\includegraphics[width=\linewidth]{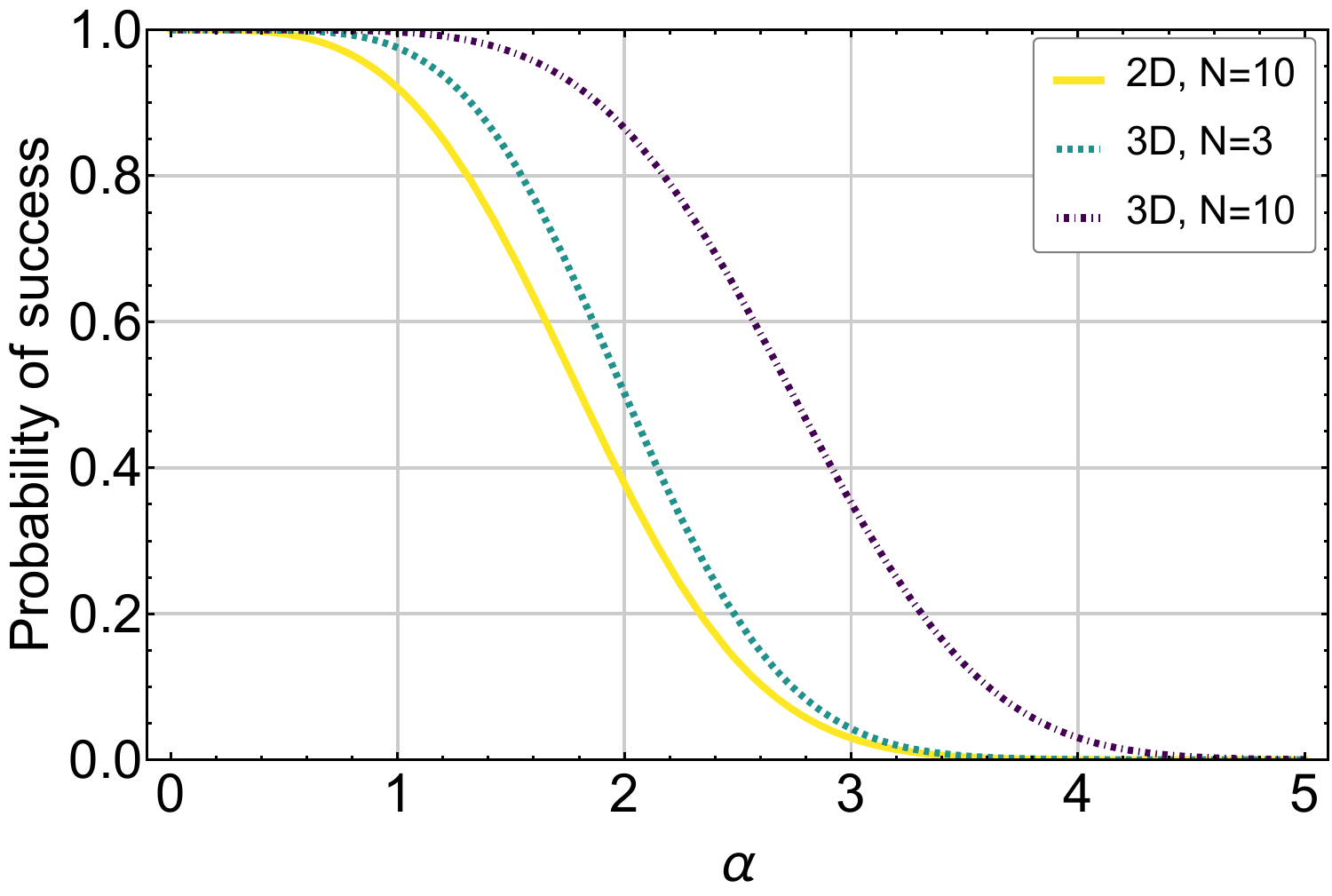}
			\caption{(b)}
			\label{fig2-2}
		\end{subfigure}
		\caption{\footnotesize{
				Teleportation with 2D and 3D channels for the input coherent state in terms of $\alpha$:
				(a) fidelity, (b) success probability, for different values ​​of $N$.}}
		\label{fig2}
	\end{figure}
	Now we will examine the results of this scheme and compare its success probability and fidelity with those of the teleportation protocol using a two-dimensional resource~\cite{8}. The plots of the success probability and fidelity as functions of the input coherent state amplitude $\alpha$ are presented in Fig.~\ref{fig2}.
	Using the 3D quantum channel of Eq.~\eqref{1} instead of the 2D one~\cite{8}, increases the dimensions of the truncated coherent states. As a result, more terms with higher photon numbers remain at the output of teleporters. Consequently, the state after the teleporters will closely resemble the input state, leading to higher teleportation fidelity. Additionally, by maintaining higher dimensions, the success probability in the teleportation process increases. The plots in Fig.~\ref{fig2} clearly demonstrate this result: by increasing the quantum channel dimension in the proposed protocol, using a smaller number of beam splitters (fewer $N$), for the same amplitude of input states, we were able to increase the fidelity and achieve a higher success probability. Notably, the number of 3 beam splitters in the proposed teleportation protocol with 3D quantum channels has even better results than 10 beam splitters in the teleportation by 2D quantum channels. Moreover, for the same $N=10$, the results are significantly improved compared to teleportation with 2D channels.\\
	
	\begin{figure*}[t]
		\centering

		\begin{subfigure}[b]{0.3\textwidth}
			\includegraphics[width=\linewidth]{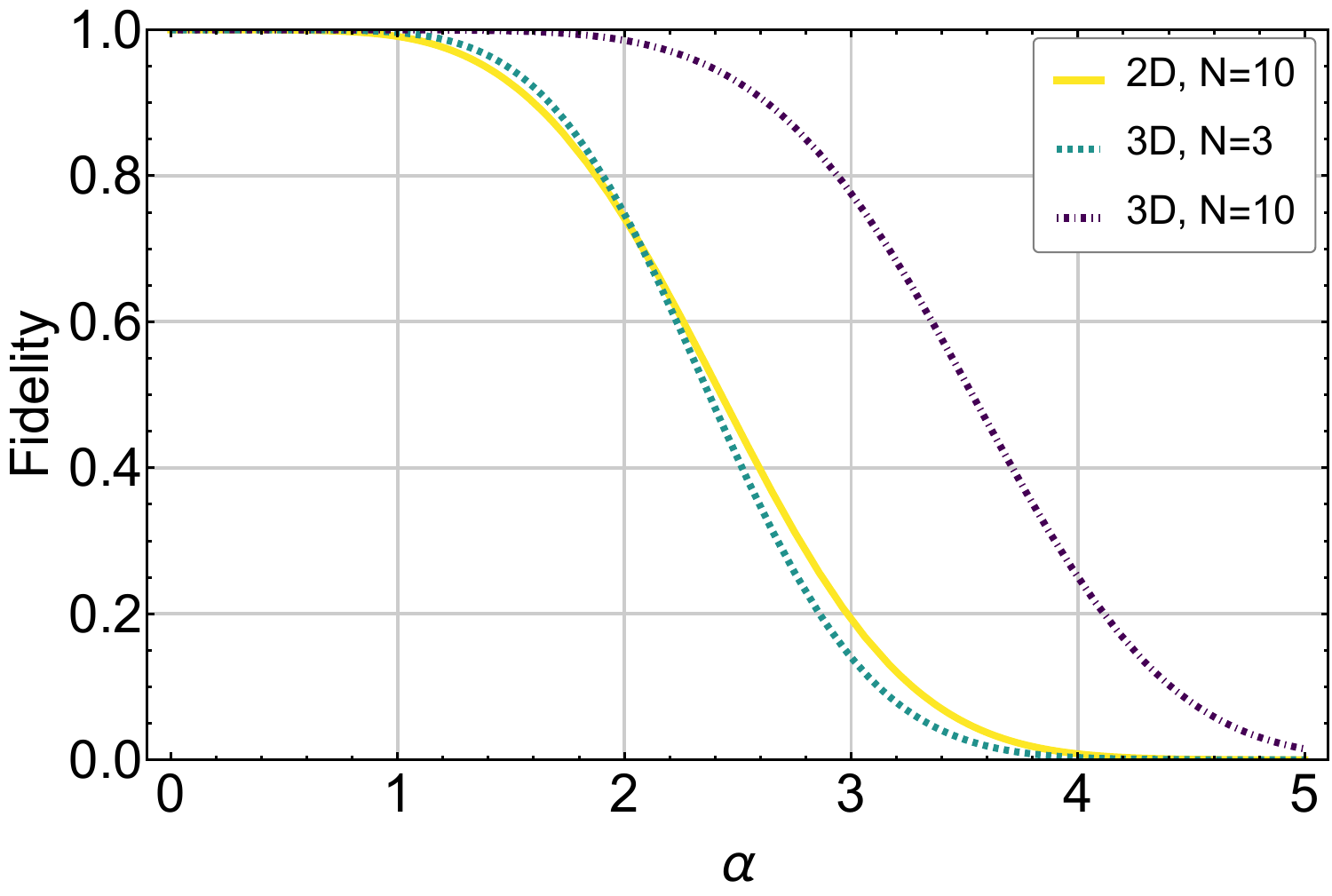}
		\end{subfigure}
		\hfill
		\begin{subfigure}[b]{0.3\textwidth}
			\includegraphics[width=\linewidth]{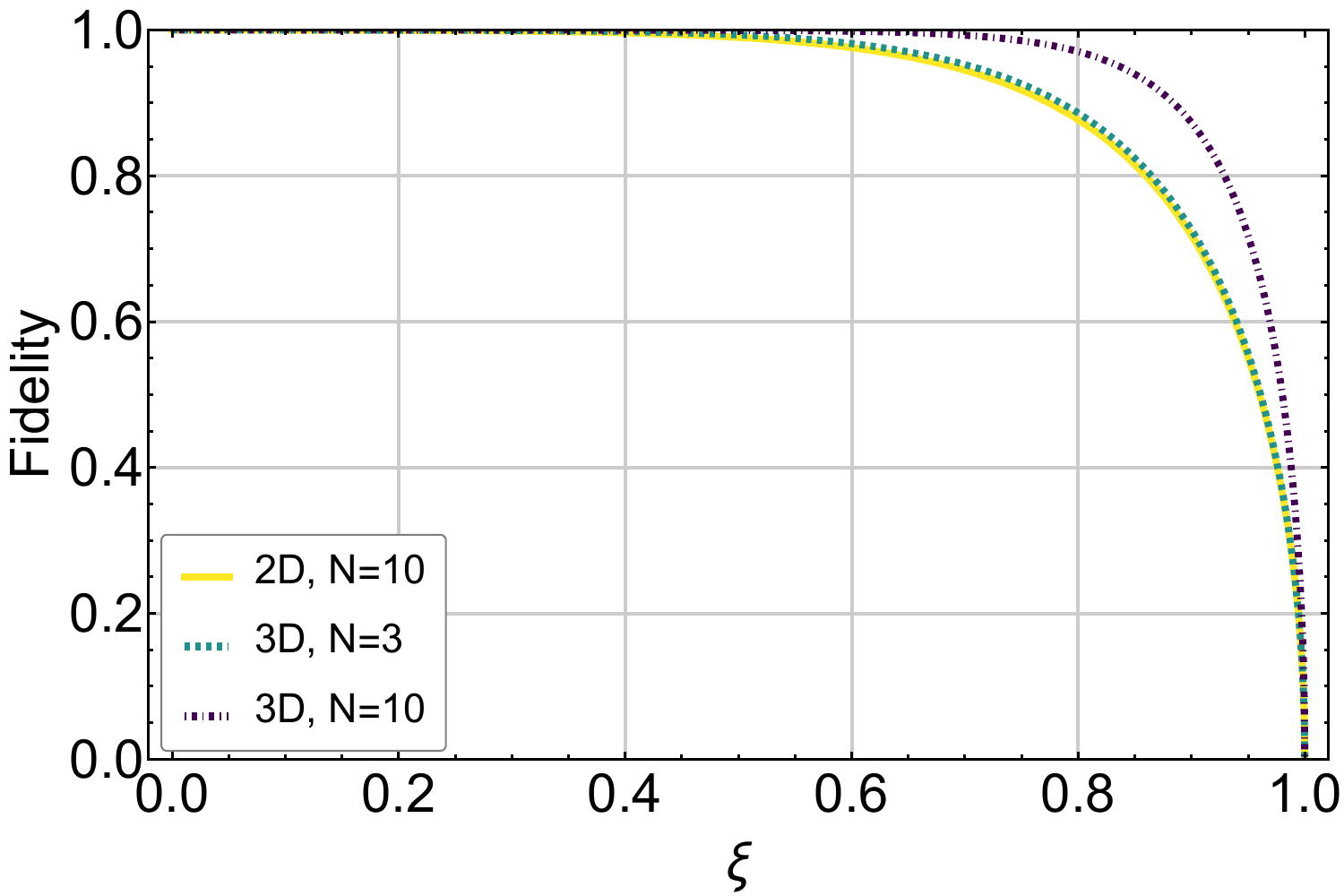}
		\end{subfigure}
		\hfill
		\begin{subfigure}[b]{0.3\textwidth}
			\includegraphics[width=\linewidth]{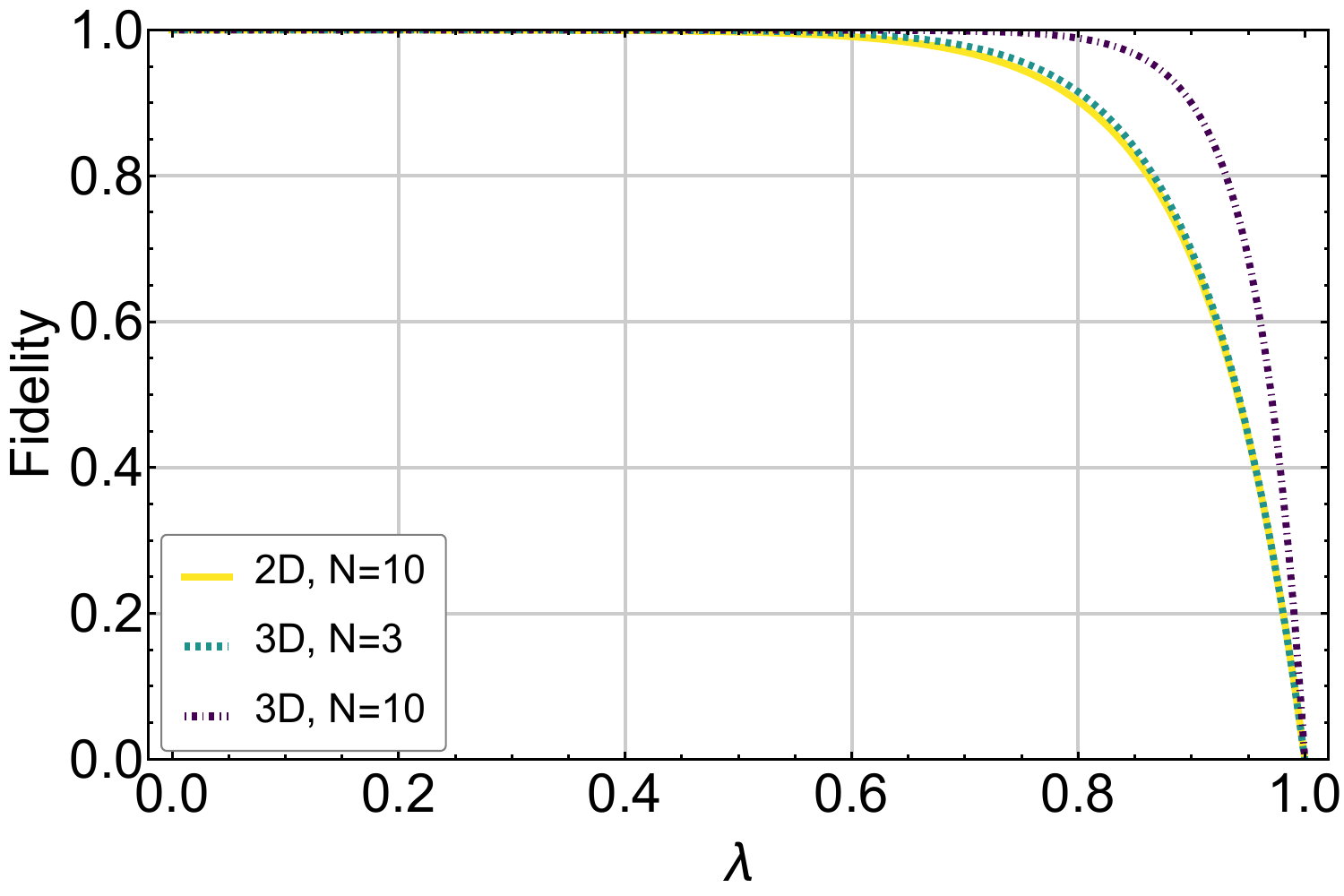}
		\end{subfigure}
		
		\vskip\baselineskip

		\begin{subfigure}[b]{0.3\textwidth}
			\includegraphics[width=\linewidth]{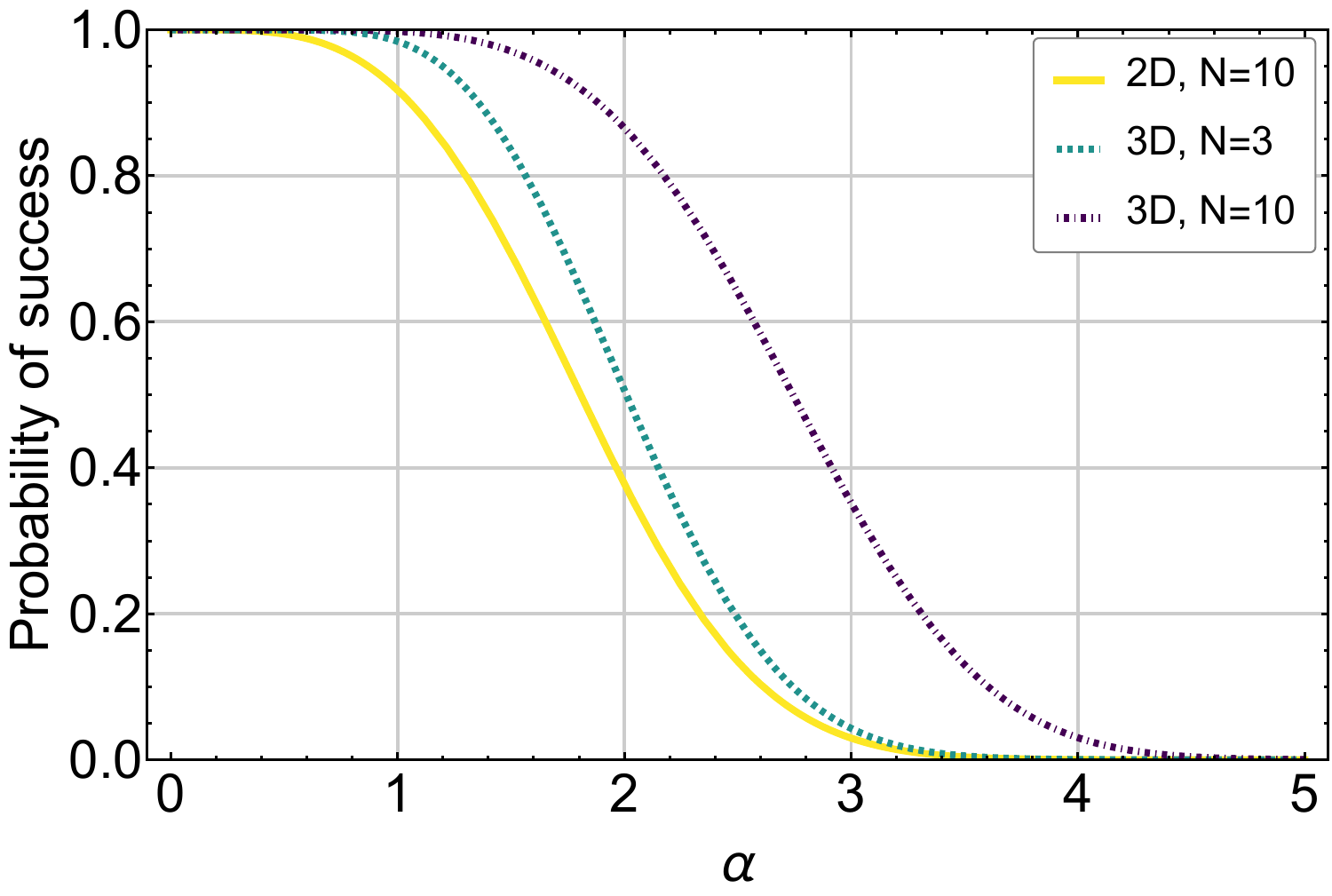}
			\caption{Schrödinger Cat State}
			\label{fig4}
		\end{subfigure}
		\hfill
		\begin{subfigure}[b]{0.3\textwidth}
			\includegraphics[width=\linewidth]{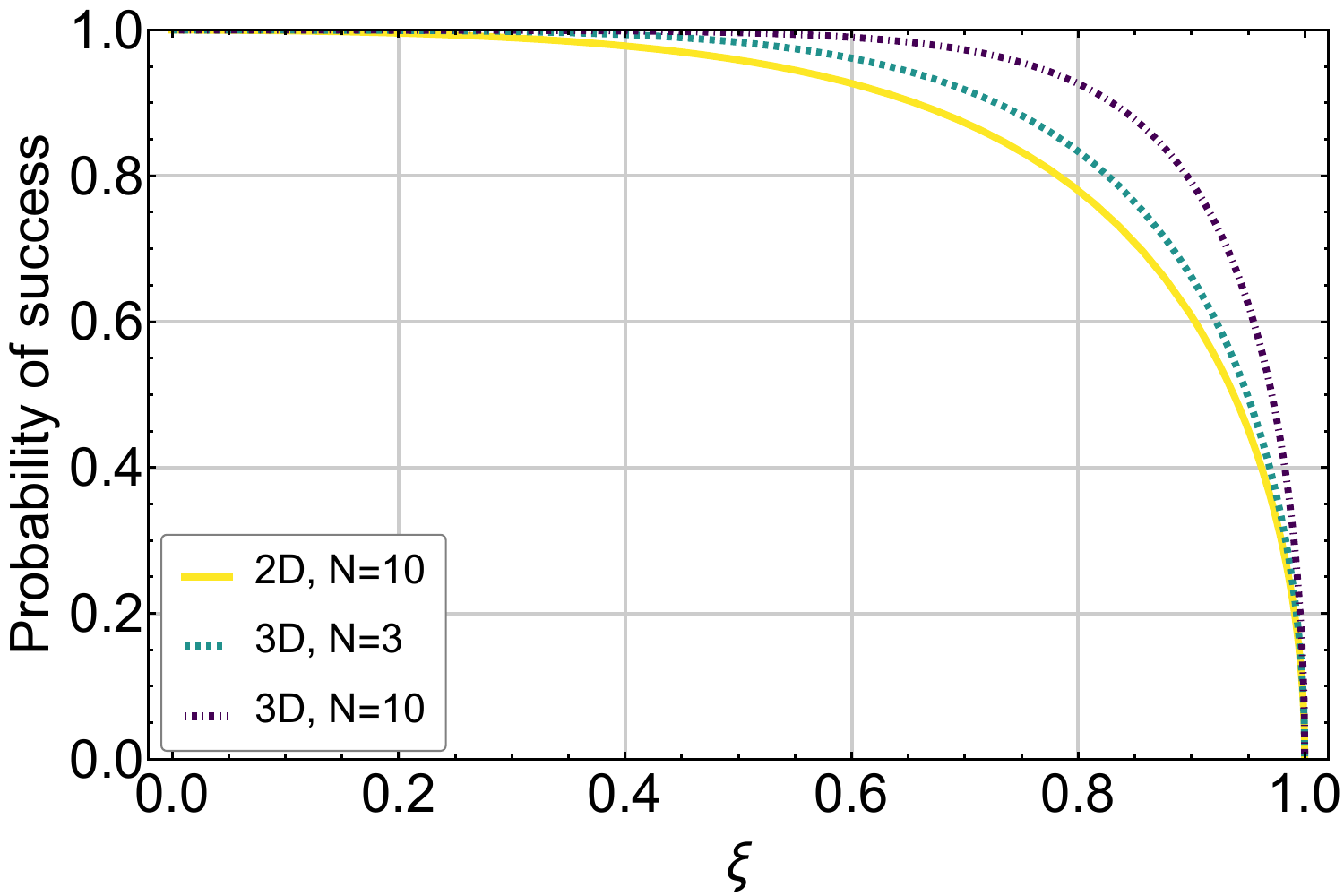}
			\caption{Vacuum Squeezed State}
			\label{fig5}
		\end{subfigure}
		\hfill
		\begin{subfigure}[b]{0.3\textwidth}
			\includegraphics[width=\linewidth]{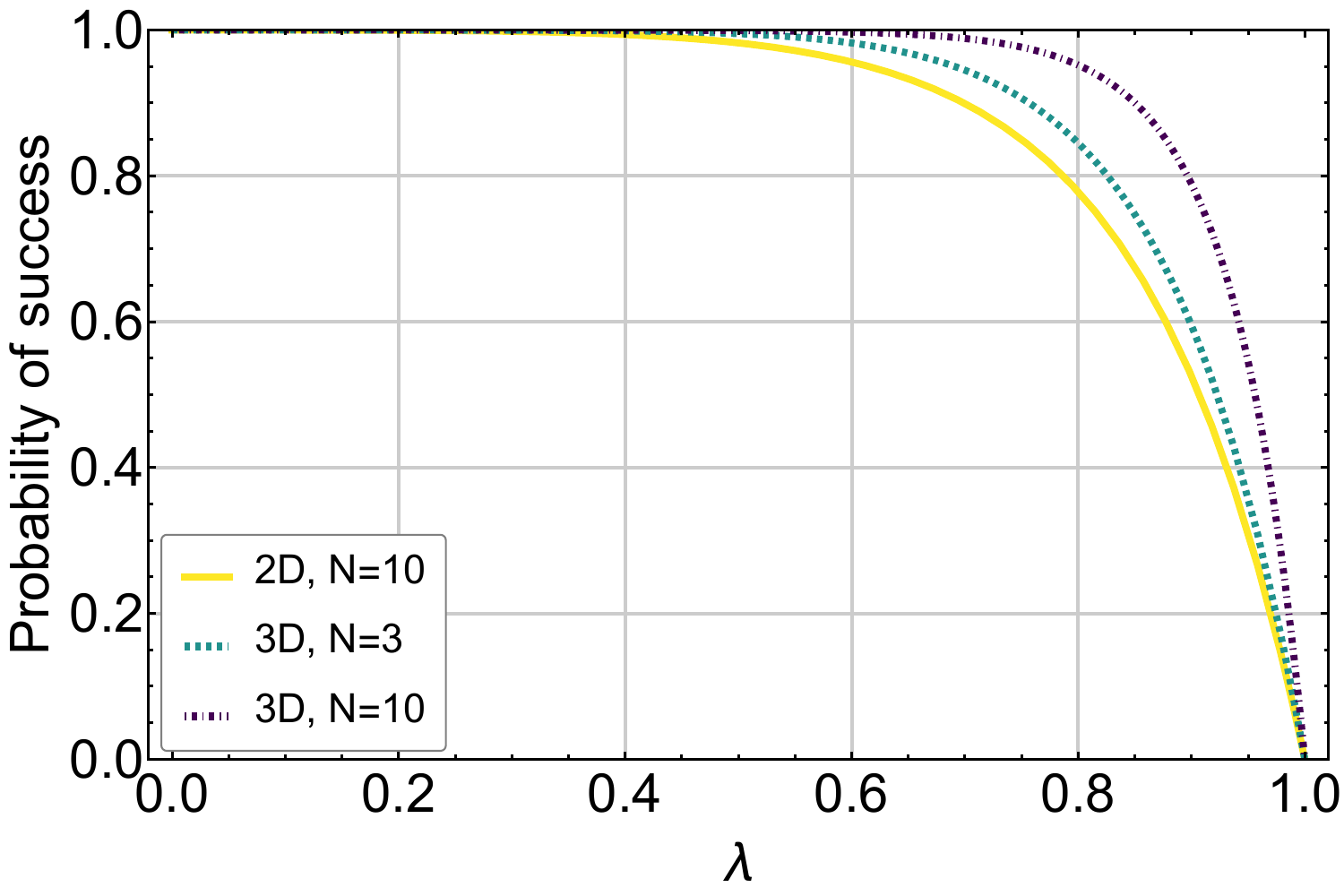}
			\caption{Two-Mode Squeezed Vacuum State}
			\label{fig6}
		\end{subfigure}
		
		\caption{\footnotesize{
				Fidelity and success probability for different input states in the interferometric teleportation scheme; (left) Schrödinger cat input state as a function of the amplitude $\alpha$, (middle) Vacuum squeezed input state as a function of the squeezing parameter $\xi $, (right)  TMSV input state as a function of the squeezing parameter $\lambda$.
		}}
		\label{fig:ideal-all}
	\end{figure*}
	\begin{figure}[h!]
		\centering
		\begin{tabular}[b]{c}
			\includegraphics[width=1\linewidth]{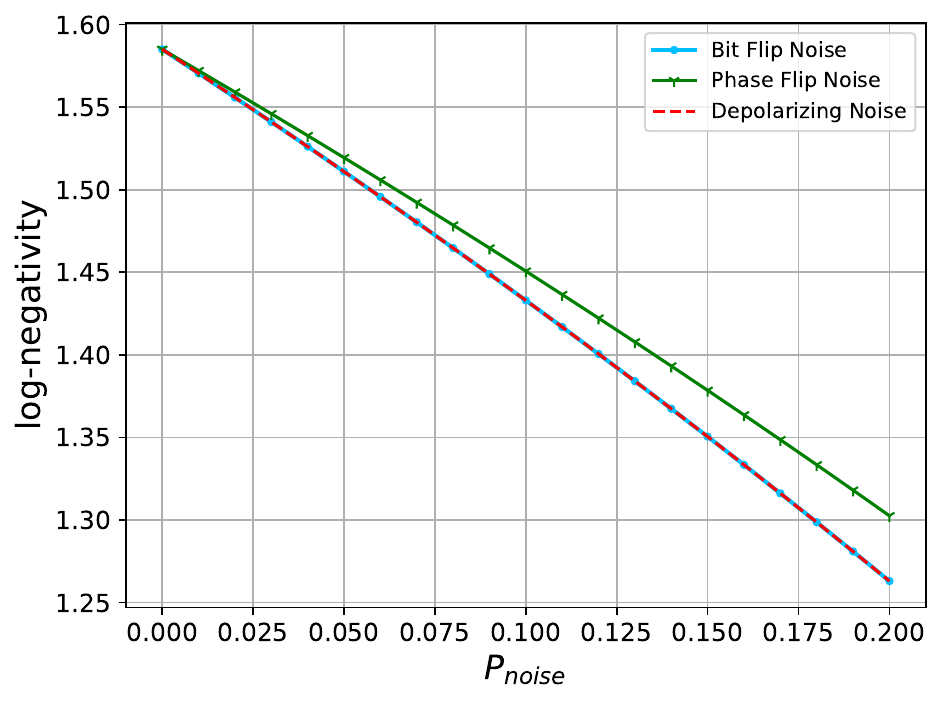}\\
		\end{tabular}
		\caption{\footnotesize{
				Logarithmic negativity of the teleportation channel as a function of the noise probability $P_{\text{noise}}$ for the bit-flip, phase-flip, and depolarizing noise models.
		}}
		\label{fig:noiseentanglement}
	\end{figure}
	By comparing the input coherent state and its final teleported output state, it is determined that the number state $\left |m\right \rangle$ is transformed through the ideal teleportation (with the 3D channels) as $ \left| m \right\rangle  \to {A_m}\left| m \right\rangle  $, where
	\begin{eqnarray}
		\label{Bm}
		{A_m} = \frac{{m!}}{{{N^m}}}N!\sum\limits_{k = \max \{ 0,m - N\} }^{\left\lfloor {{\raise0.7ex\hbox{$m$} \!\mathord{\left/
						{\vphantom {m 2}}\right.\kern-\nulldelimiterspace}
					\!\lower0.7ex\hbox{$2$}}} \right\rfloor } {\frac{1}{{k!(m - 2k)!(N - m + k)!}}} \, .
	\end{eqnarray}
	Therefore, the protocol can be extended to teleportation of an arbitrary state by expanding other input states in terms of number states and using the above transformation. Using this approach, the success probability and fidelity of proposed protocol are plotted in Fig.~\ref{fig:ideal-all}, for three different input states: Schrödinger's cat state ${\left| \psi  \right\rangle _{\text{cat}}} = \frac{{\left| \alpha  \right\rangle  + \left| { - \alpha } \right\rangle }}{{\sqrt {2 + 2\exp ( - 2{\alpha ^2})} }}$, vacuum squeezed state $|\xi\rangle = (1 - |\xi|^2)^{1/4} \sum\limits_{n=0}^{\infty} \frac{(-1)^n \sqrt{(2n)!}}{2^n n!}\,\xi^n\, |2n\rangle$ with squeezing parameter $\xi$, and TMSV state $\label{tmsv} {\left| \text{TMSV}  \right\rangle } = \sqrt {1 - {{\left| \lambda  \right|}^2}} \sum\limits_{n = 0}^\infty  {{\lambda ^n}\left| {n,n} \right\rangle }$ with squeezing parameter $\lambda$ (teleportation of one mode). This figure clearly demonstrates that for all the input states, the proposed 3D scheme achives fidelity and success probability comparable, or even exceeding, that of 2D approach, despite using fewer beam-splitters (3 vs. 10). Also when $N=10$ outcomes improve substantially in comparison with teleportation based on 2D channels. \\
	\begin{figure*}[t]
		\centering

		\begin{subfigure}[b]{0.3\textwidth}
			\includegraphics[width=\linewidth]{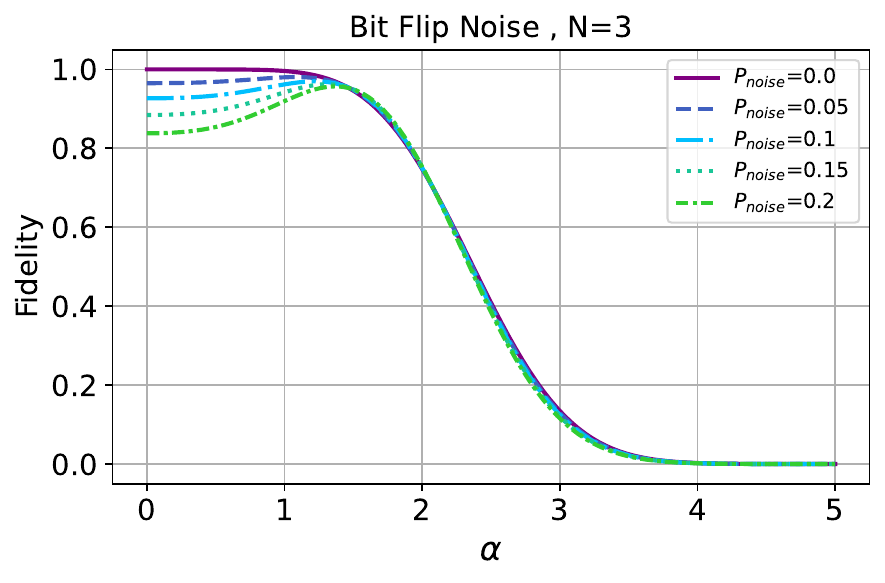}
		\end{subfigure}
		\hfill
		\begin{subfigure}[b]{0.3\textwidth}
			\includegraphics[width=\linewidth]{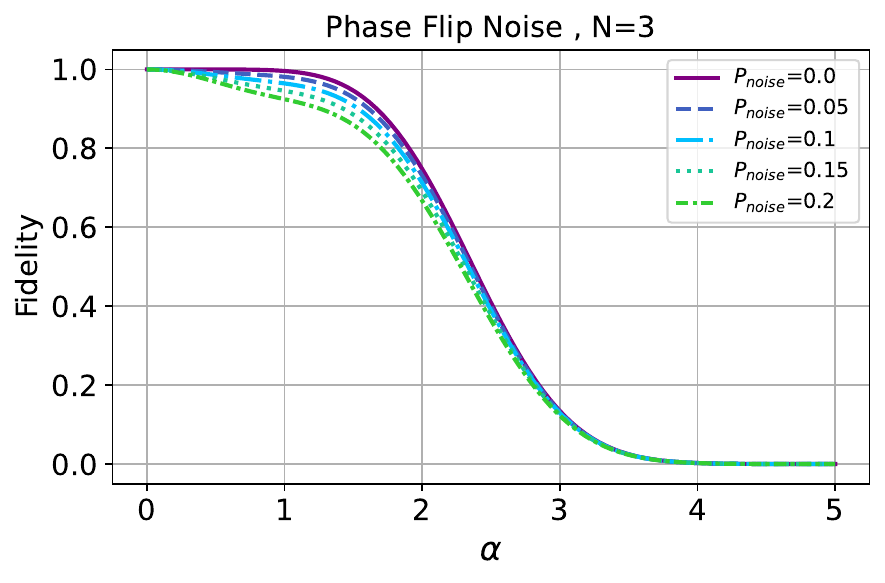}
		\end{subfigure}
		\hfill
		\begin{subfigure}[b]{0.3\textwidth}
			\includegraphics[width=\linewidth]{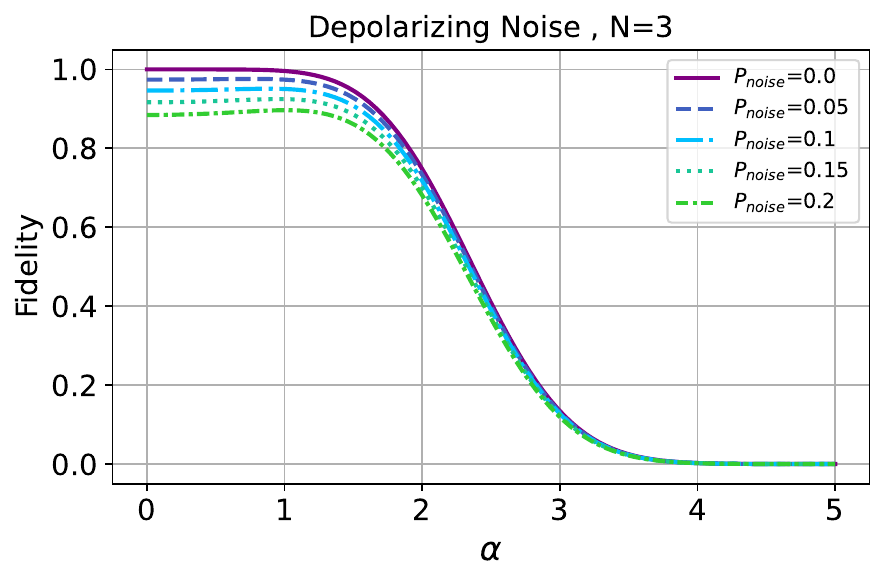}
		\end{subfigure}
		
		\vskip\baselineskip

		\begin{subfigure}[b]{0.3\textwidth}
			\includegraphics[width=\linewidth]{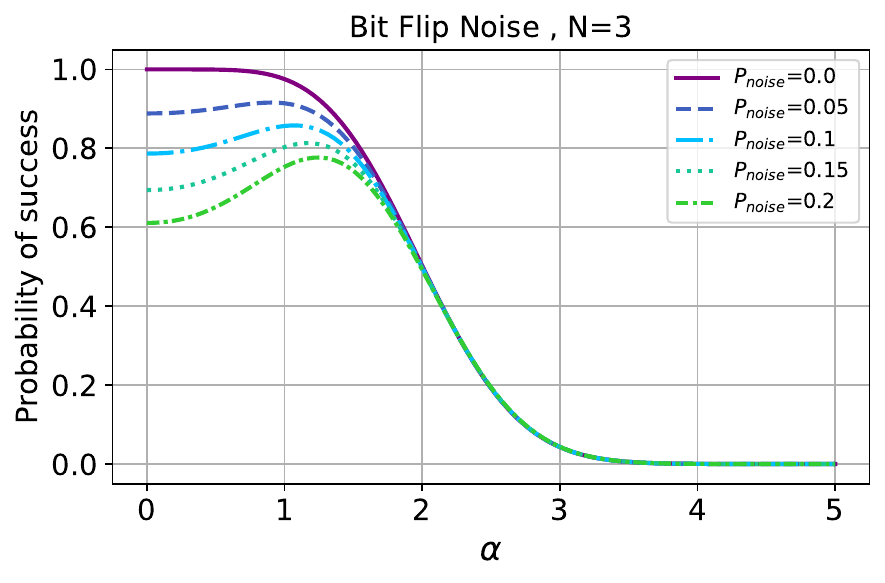}
		\end{subfigure}
		\hfill
		\begin{subfigure}[b]{0.3\textwidth}
			\includegraphics[width=\linewidth]{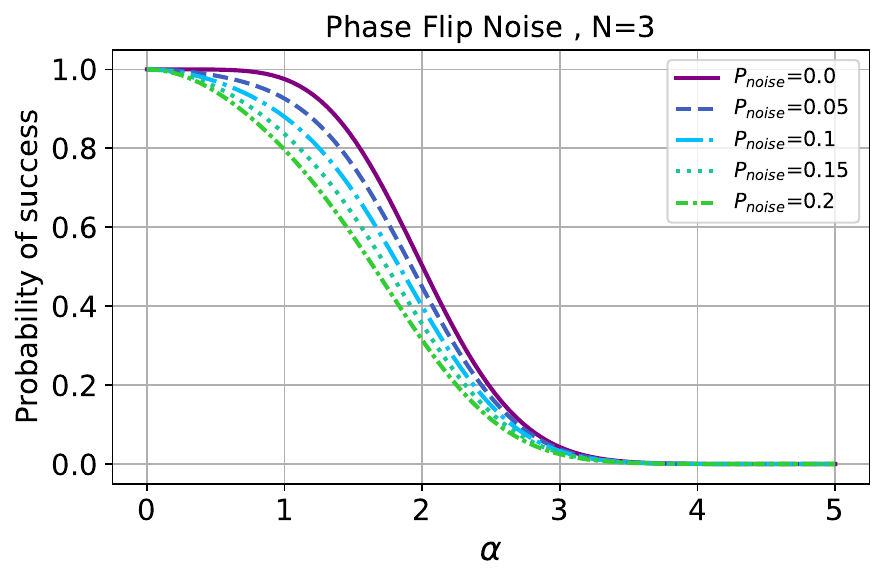}
		\end{subfigure}
		\hfill
		\begin{subfigure}[b]{0.3\textwidth}
			\includegraphics[width=\linewidth]{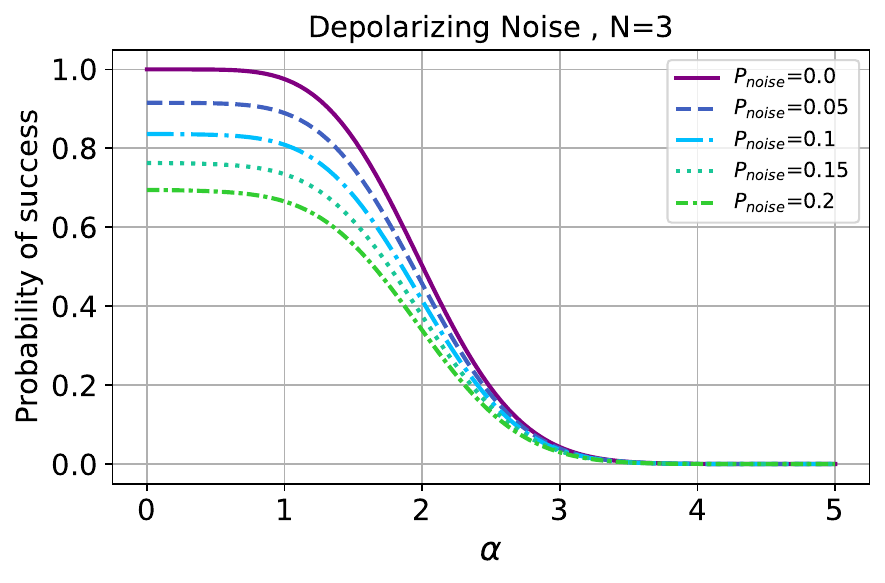}
		\end{subfigure}
		
		\caption{\footnotesize{
				Fidelity and success probability for the input coherent state under different quantum noises applied to the teleporters for $N=3$ and different values ​​of the noise probability $P_{\text{noise}}$; (left) bit-flip noise, (middle) phase-flip noise, (right) depolarizing noise.
		}}
		\label{fig:noisy-all}
	\end{figure*}

	\label{section3}
	\section{Impact of the noise on a multimode interference-based teleportation scheme with three-dimensional quantum channels}
	
	Since the maximally entangled state of Eq.~\ref{1}, is affected by noise and environmental interactions, under realistic conditions, its degree of entanglement is reduced. Therefore, one cannot expect perfect teleportation for states with up to three photons. As a result, under realistic conditions, Eq.~\ref{3} no longer holds. In this section, we examine the proposed teleportation scheme under realistic noisy conditions by applying various types of noise to the teleporter arms.\\
	In~\cite{14} it is demonstrated that standard teleportation using an arbitrary mixed-state resource is equivalent to a generalized depolarizing channel, whose probabilities are determined by the maximally entangled components of the resource. Thus, any non-maximal entangled teleporter can be considered as a noisy quantum channel.\\
	To investigate the effect of noise on the fidelity and the success probability of the teleportation, we consider the coherent state as the input state. As mentioned, the input state first enters an \textit{N}-splitter and is converted to a state with reduced amplitude of Eq.~\eqref{3}. Therefore, the density matrix of the input state to each teleporter is $\rho_{\text{in}}^{\text{tel}}  = \left| {\frac{\alpha }{{\sqrt N }}} \right\rangle \left\langle {\frac{\alpha }{{\sqrt N }}} \right|$, which, under the noisy channel becomes
	\begin{eqnarray}
		\label{11}
		\rho_{\text{out}}^{\text{tel}} = \sum\limits_i {{K_i}\rho_{\text{in}}^{\text{tel}} {K_i}^\dag } \, .
	\end{eqnarray}
	In this relation, ${K_i}$ are the Kraus operators related to the channel and its associated noise, and satisfy normalization condition $\sum_i K_i^{\dagger} K_i = \mathbb{1}$. The number and type of these operators vary for different types of noisy channels. Ref~\cite{13} lists the Kraus operators related to different noise types that may affect a qutrit. In this paper, three quantum noises are considered: Bit-Flip noise, Phase-Flip noise, and Depolarizing noise.
	The bit-flip noise acts as a cyclic permutation on the computational basis states, causing $\left |0\right \rangle$ to transition to $\left |1\right \rangle$ (or $\left |2\right \rangle$) , $\left |1\right \rangle$ to $\left |2\right \rangle$ (or $\left |0\right \rangle$), and $\left |2\right \rangle$ to $\left |0\right \rangle$ (or $\left |1\right \rangle$) with a certain probability. This process introduces computational errors by randomly swapping the logical values of the qutrit, thereby corrupting the information encoded in the state. 
	In contrast, phase-flip noise does not alter the populations of the basis states but instead manipulates their relative phases. By introducing a phase shift of $-1$ to the $\left |1\right \rangle$ and $\left |2\right \rangle$ components, this noise disrupts the delicate coherence of the superposition and entanglement, which is crucial for the interference effects underlying quantum teleportation. Finally, depolarizing noise is the most disruptive of the three models. It randomizes the state by, with a certain probability, replacing it with a maximally mixed state. This process erodes both the population and coherence of the quantum information, effectively diminishing the "quantumness" of the state. These qualitative transformations, formally captured by the Kraus operators detailed in Appendix~\ref{A}, are the underlying physical mechanisms that lead to the degradation in teleportation fidelity.
	\\
	\begin{figure*}[t]
		\centering

		\begin{subfigure}[b]{0.3\textwidth}
			\includegraphics[width=\linewidth]{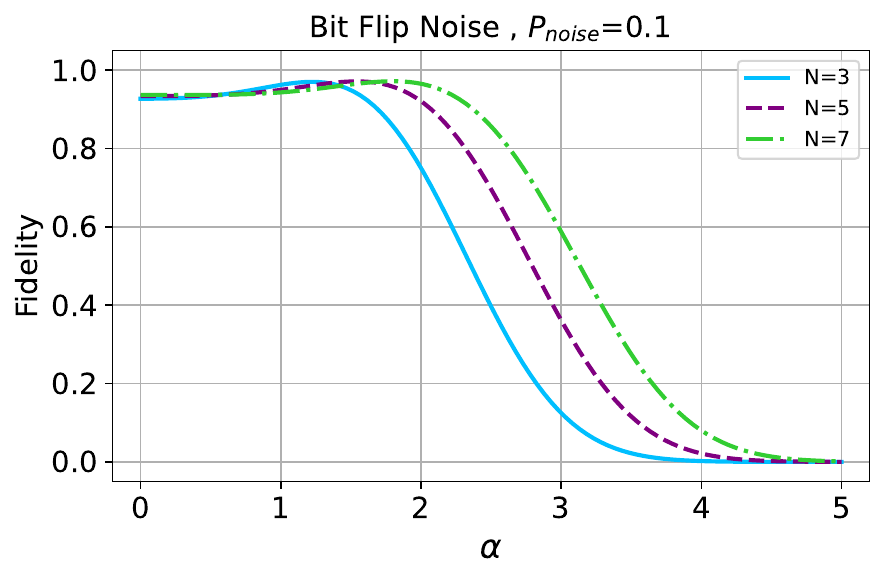}
		\end{subfigure}
		\hfill
		\begin{subfigure}[b]{0.3\textwidth}
			\includegraphics[width=\linewidth]{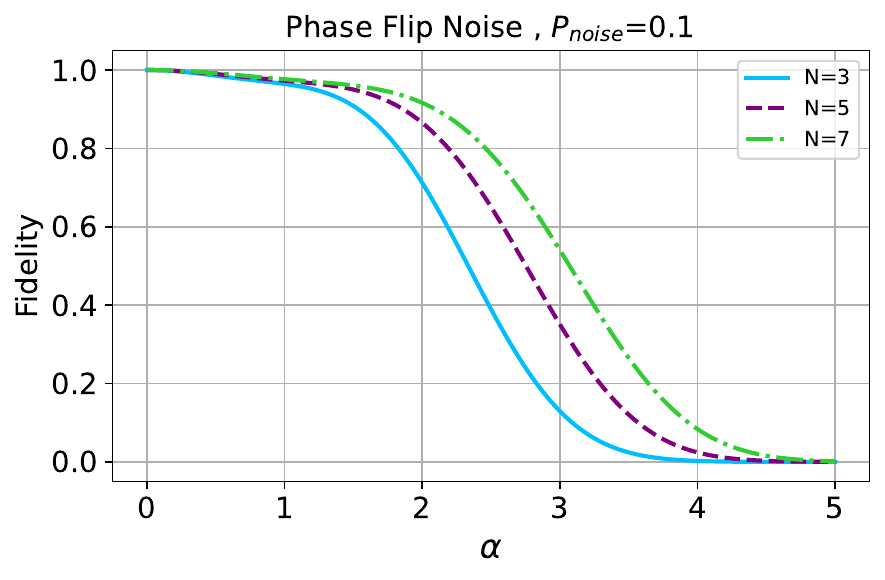}
		\end{subfigure}
		\hfill
		\begin{subfigure}[b]{0.3\textwidth}
			\includegraphics[width=\linewidth]{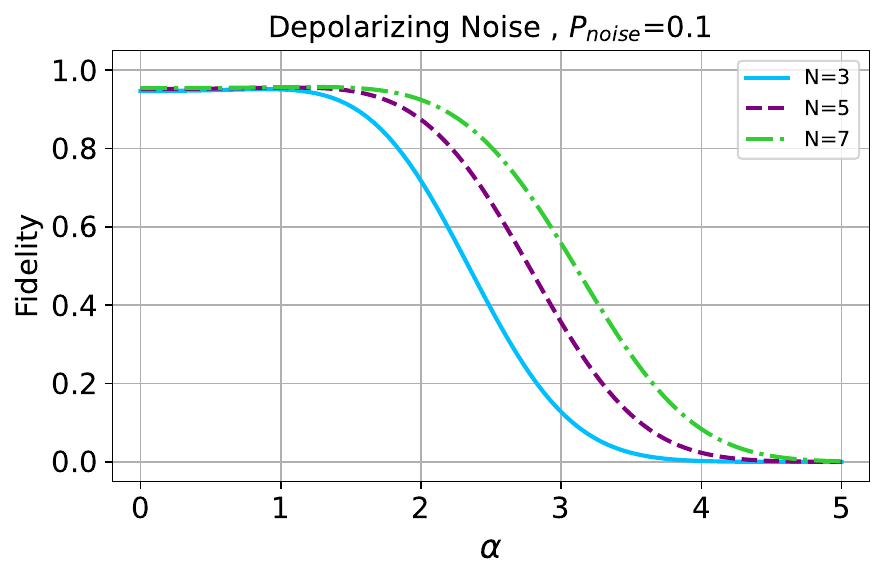}
		\end{subfigure}
		
		\vskip\baselineskip

		\begin{subfigure}[b]{0.3\textwidth}
			\includegraphics[width=\linewidth]{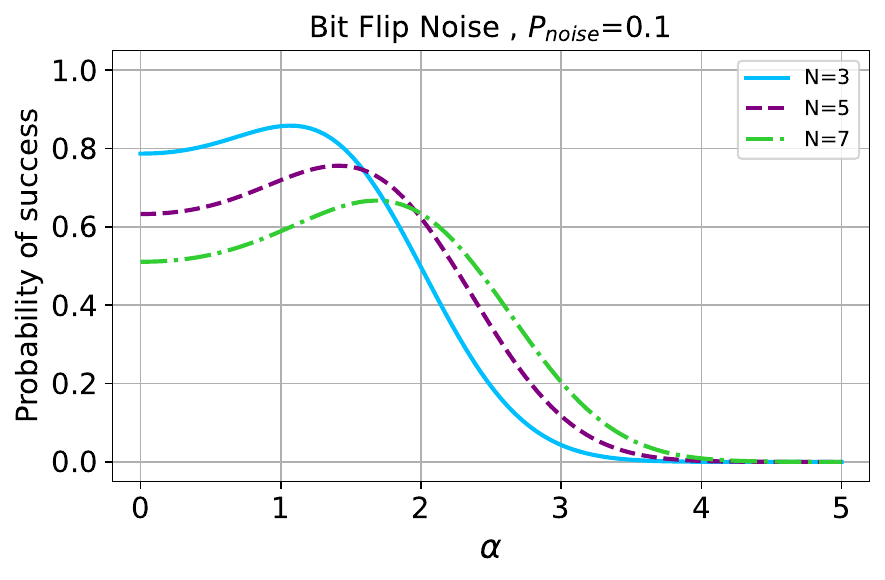}
		\end{subfigure}
		\hfill
		\begin{subfigure}[b]{0.3\textwidth}
			\includegraphics[width=\linewidth]{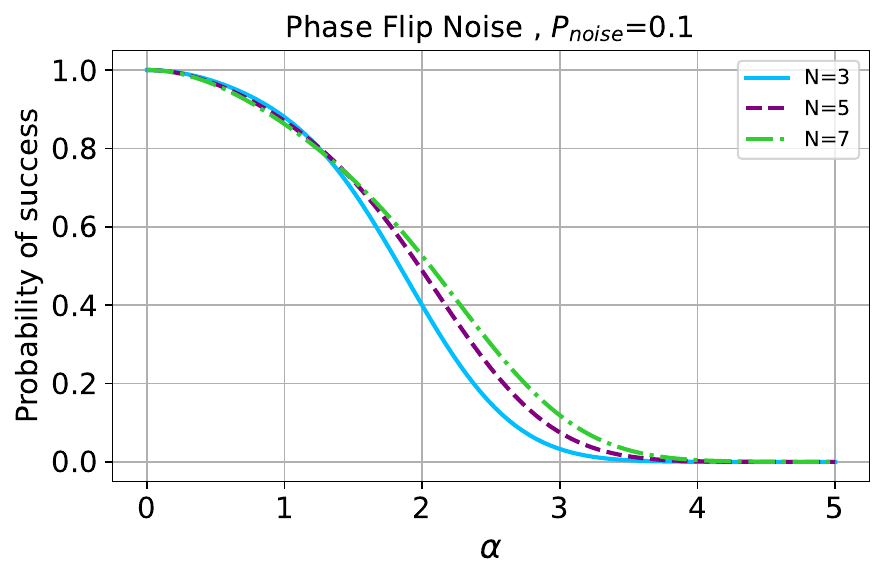}
		\end{subfigure}
		\hfill
		\begin{subfigure}[b]{0.3\textwidth}
			\includegraphics[width=\linewidth]{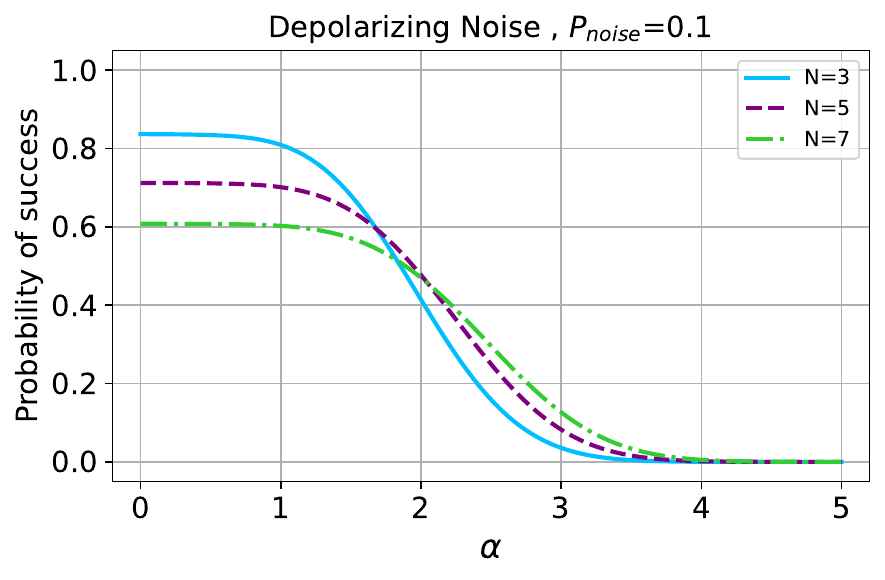}
		\end{subfigure}
		
		\caption{\footnotesize{
				Fidelity and success probability for the input coherent state under different quantum noises applied to the teleporters for different number of beam-splitters $N$; (left) bit-flip noise, (middle) phase-flip noise, (right) depolarizing noise.
		}}
		\label{fig:noisy-differn-all}
	\end{figure*}	
	\begin{figure*}[t]
		\centering

		\begin{subfigure}[b]{0.3\textwidth}
			\includegraphics[width=\linewidth]{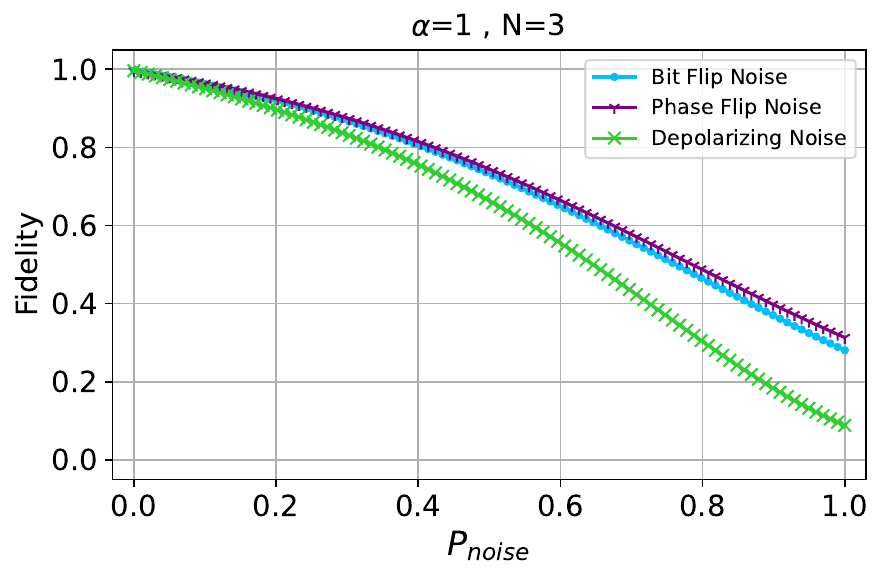}
		\end{subfigure}
		\hfill
		\begin{subfigure}[b]{0.3\textwidth}
			\includegraphics[width=\linewidth]{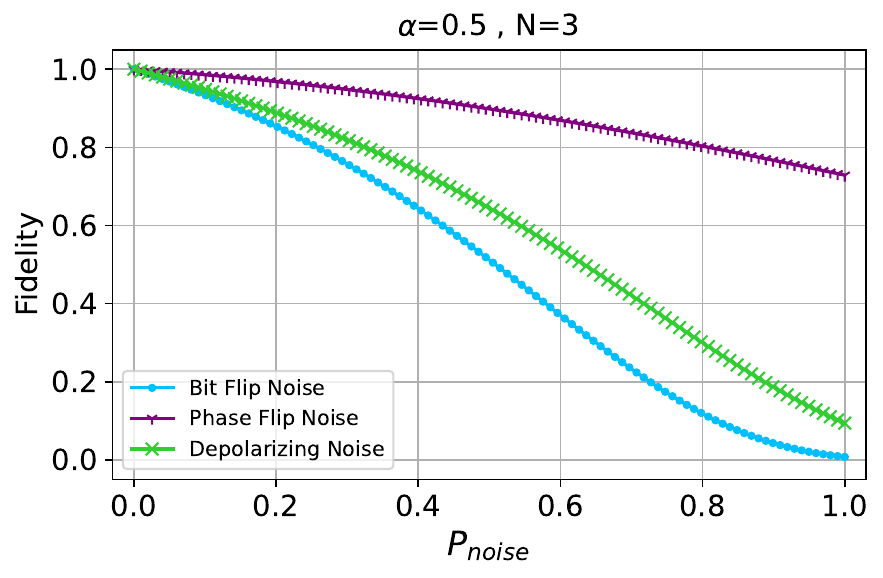}
		\end{subfigure}
		\hfill
		\begin{subfigure}[b]{0.3\textwidth}
			\includegraphics[width=\linewidth]{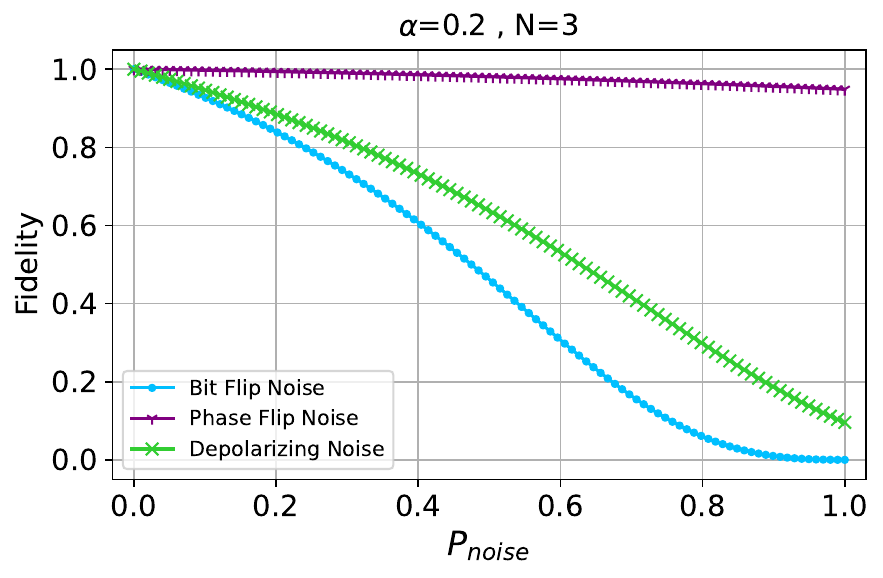}
		\end{subfigure}
		
		\vskip\baselineskip

		\begin{subfigure}[b]{0.3\textwidth}
			\includegraphics[width=\linewidth]{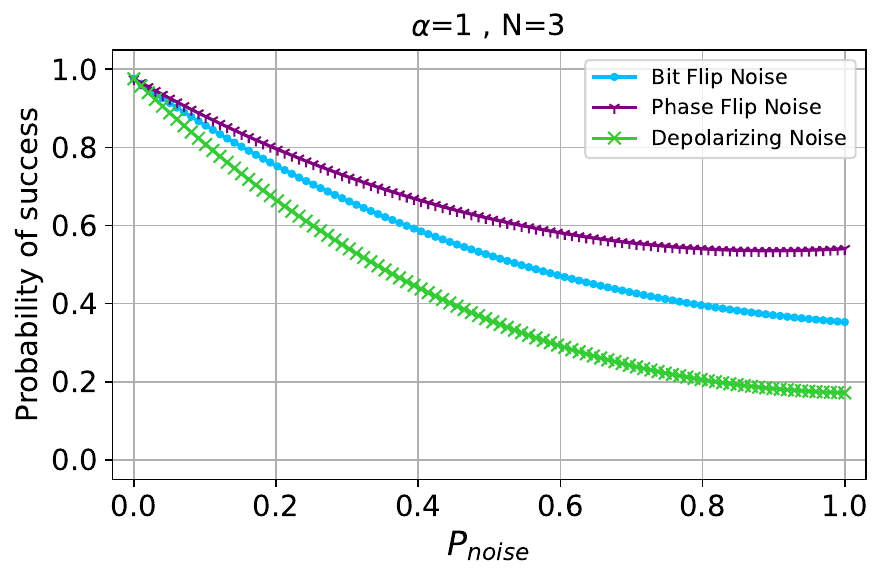}
		\end{subfigure}
		\hfill
		\begin{subfigure}[b]{0.3\textwidth}
			\includegraphics[width=\linewidth]{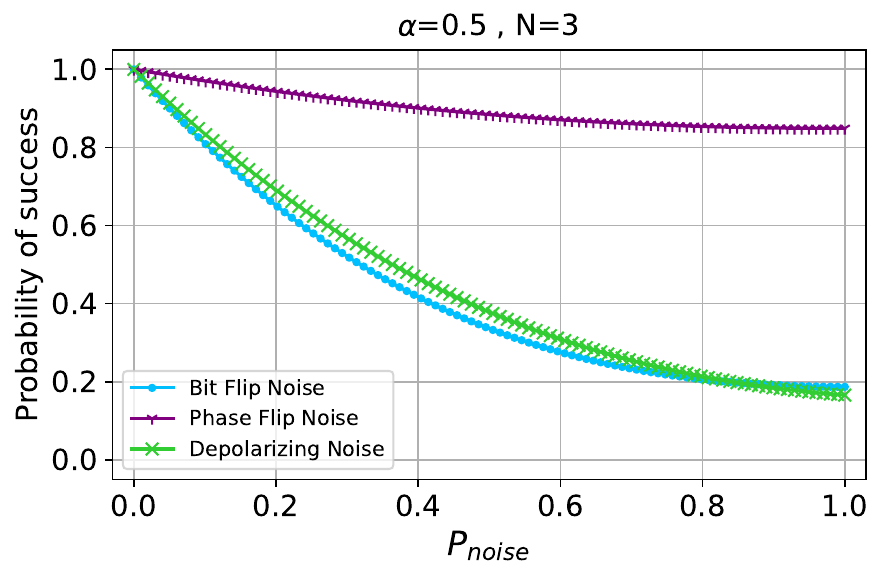}
		\end{subfigure}
		\hfill
		\begin{subfigure}[b]{0.3\textwidth}
			\includegraphics[width=\linewidth]{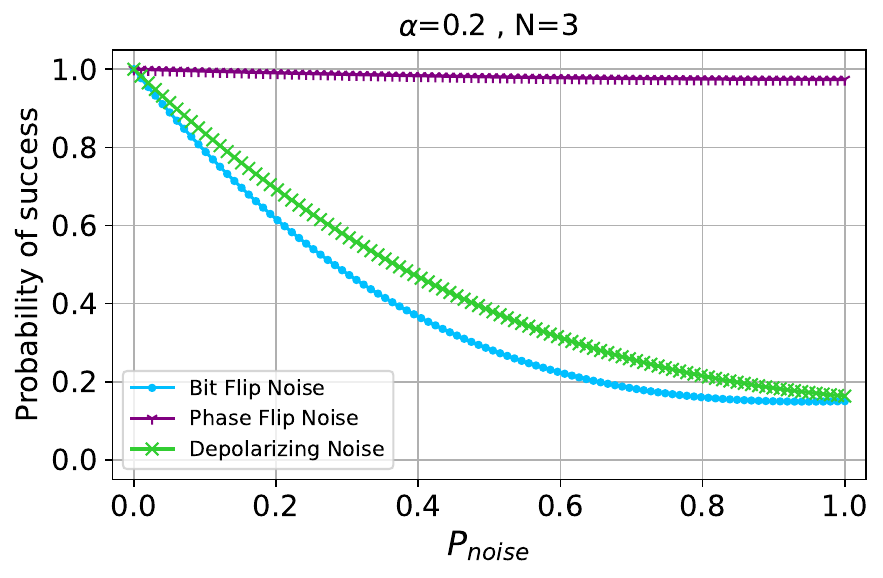}
		\end{subfigure}
		
		\caption{\footnotesize{
				Fidelity and success probability for the input coherent state under different quantum noises applied to the teleporters for $N=3$ and three different values ​​of the coherent state amplitude $\alpha$; (left) bit-flip noise, (middle) phase-flip noise, (right) depolarizing noise.
		}}
		\label{fig:all noise}
	\end{figure*}	
	
	In examining the effects of noise, it is assumed that $P_{\text{noise}}$ is the probability of the channel noise affecting the input state. Therefore $P_{\text{noise}}=0$ represents ideal, noiseless conditions, while $P_{\text{noise}}=1$ describes completely noisy conditions.\\
	\textcolor{black}{
		Building upon the equivalence established in~\cite{14}, where standard teleportation with an arbitrary mixed entangled resource is shown to be equivalent to a generalized depolarizing channel whose Kraus probabilities are determined by the weights of the maximally entangled components of the resource state, we further quantify how the entanglement of the teleportation channel depends on the noise probability $P_{\text{noise}}$.
		Therefore, instead of explicitly reconstructing the non-maximally entangled shared state and re-deriving the full qutrit teleportation protocol (including generalized Bell measurements and conditional local corrections), we adopt the equivalent and widely used approach in which the non-ideal teleportation process is modeled directly as a noisy quantum channel described by Kraus operators. In this framework, the squared amplitudes associated with the Kraus operators correspond to the contributions of the maximally entangled (generalized Bell) states in the expansion of the shared resource. Consequently, the entanglement of the effective channel can be evaluated directly in terms of these coefficients. To this end, we compute the logarithmic negativity of the corresponding shared state as a function of the noise probability for three noise models - bit-flip, phase-flip, and depolarizing noise - within the interval $0\le P_{\text{noise}} \le 0.2$. The results are presented in Fig.~\ref{fig:noiseentanglement}. As observed, in all three cases the entanglement decreases with increasing noise probability in the considered range. Moreover, the degradation of entanglement for the bit-flip and depolarizing channels is identical and more pronounced than that of the phase-flip channel. It is important to note that, although the reduction of entanglement for the bit-flip and depolarizing channels is identical within the considered range, the resulting mixed states of the teleportation channel are not the same. Consequently, despite exhibiting similar entanglement degradation, these two noise models can lead to different effects on the teleportation of quantum states.}
	\\
	By applying the Kraus operators to the input state of the teleporters, Eq.~\eqref{11}, the output state of the teleporters is obtained. Finally, as described in the proposed teleportation scheme (Fig.~\ref{fig:nsplitter}), these output states are re-entered into an \textit{N}-splitter, where, through projecting all the output modes except one onto the state $\left| 0 \right\rangle $, the final output state is obtained. It can be shown that the final output state is
	\begin{equation}
		{\rho _{\text{out}}} =\frac{1}{P_\text{s}^{\text{noisy}}} \sum\limits_{0 \le p,q \le 2N} {{D_{pq}}\left| p \right\rangle \left\langle q \right|} \, ,
	\end{equation}
	whose matrix elements in the Fock basis are as follows:
	\begin{equation}
		\label{13}
		D_{pq} = 
		\frac{\sqrt{p!\,q!}}{\sqrt{N^{p+q}}}
		\mathop{\mathlarger{\mathlarger{\sum}}}\limits_{\substack{
				n_1,\ldots,n_N\\
				m_1,\ldots,m_N\\
				n_1+\cdots+n_N = p\\
				m_1+\cdots+m_N = q
		}}
		\prod\limits_{i = 1}^N {\frac{{{{(\rho _{\text{out}}^{\text{tel}})}_{{n_i}{m_i}}}}}{{\sqrt {{n_i}!\,\,{m_i}!} }}} \,.
	\end{equation}
	In this relation, ${{{(\rho _{\text{out}}^{\text{tel}})}_{{n_i}{m_i}}}}$ are the matrix elements of $ \rho_{\text{out}}^{\text{tel}}$ of the Eq.~\eqref{11}. 
	The success probability $P_{\text{s}}^{\text{noisy}}$ obtained from the normalization condition $tr({\rho _{\text{out}}})=1$ and the teleportation fidility is $\left\langle \alpha  \right|\rho _{\text{out}}\left| \alpha  \right\rangle $ which investigated for the different types of quantum noises. The results are shown in Figs.~\ref{fig:noisy-all},\ref{fig:noisy-differn-all},\ref{fig:all noise}.\\
	
	Fig.~\ref{fig:noisy-all} shows the fidelity and the success probability of three types of noise, in terms of the amplitude of the input coherent state $(\alpha)$, for different levels ​​of the channel noise $P_{\text{noise}}$. As the ideal case (discussed in Sec.~\ref{section2}) showed acceptable fidelity and success probability for $N=3$, the same number of teleporters has been employed in Fig.~\ref{fig:noisy-all} for the numerical simulations.\\
	The results indicate that, in general, the presence of noise reduces both the fidelity and the success probability in quantum teleportation. In the presence of phase-flip and depolarizing noise, as the amplitude of the coherent state $\alpha$ increases, both the fidelity and the success probability decrease overall for all noise strengths. However, in the case of bit-flip noise, it is observed that both the fidelity and the success probability initially increase for small values of the amplitude $\alpha$ at any nonzero noise probability ${P_{\text{noise}}}\neq0$, and then decrease as the amplitude grows. In other words, a peak appears in the corresponding plots. This behavior arises from the specific nature of the bit-flip noise.
	For small amplitudes $\alpha < 1$, the dominant term in the input state of the teleporter arms, $\left| {\frac{\alpha}{\sqrt{N}}} \right\rangle$, is the vacuum state $\left| 0 \right\rangle$. Therefore when bit-flip noise is applied, the state $\left| 0 \right\rangle$ can be transformed into $\left| 1 \right\rangle$ or $\left| 2 \right\rangle$ (see Appendix~\ref{A}), meaning that the input state is almost completely altered. Therefore, a significant reduction in both the fidelity and the success probability is observed in the presence of this type of noise for small amplitudes. But as the amplitude increases to about $1.5$, the probability of presence states $\left| 1 \right\rangle $ and $\left| 2 \right\rangle $ in the input state increases and approaches to the probability of presence state $\left| 0 \right\rangle $, and with the bit-flip noise there is a probability that the noisy output state of a non-ideal teleporter becomes very similar to its input state. Therefore, we see an increase in both fidelity and success probability in this range of the input amplitude $\alpha$. Even for $\alpha\simeq1.5$, the noisy state is almost the same as the ideal state. But as the amplitude of the coherent input state increases further, the dominant factor in reducing fidelity and success probability is the finite dimension of the teleporters (three in our scheme), and at all noise strengths, increasing the amplitude of $\alpha$ is accompanied by a sharp decrease in fidelity and success probability.\\
	
	In another analysis, the fidelity and success probability are plotted as functions of the amplitude $\alpha$ for a fixed noise probability of $P_{\text{noise}}=0.1$ and different numbers of teleporters $N$ (see Fig.~\ref{fig:noisy-differn-all}).
	In the ideal noise-free case, increasing the number of beam-splitters improves the teleportation fidelity. In the presence of phase-flip and depolarizing noise, a similar trend is observed. However, in the teleportation fidelity diagram for bit-flip noise, the opposite behavior occurs for amplitude values close to one, that is, increasing $N$ leads to a slight decrease in fidelity.
	This behavior can be explained as follows: for small amplitudes, with increasing $N$, the contribution of the vacuum term $\left| 0 \right\rangle $ in the input state of the teleporters becomes larger, which is transformed into a completely different state ($\left| 1 \right\rangle $ or $\left| 2 \right\rangle $) under the influence of the bit-flip noise. However, in this figure, due to a relatively low noise level (10\%), the reduction in fidelity is not substantial. Furthermore, the success probability diagrams in Fig.~\ref{fig:noisy-differn-all} show that with increasing $N$, at lower values ​​of $\alpha$, the probability of success decreases (of course, in phase-flip noise, this decrease is insignificant) and at larger values ​​of $\alpha$, it increases. These results indicate that, under realistic noisy conditions and for small input amplitudes $\alpha$, by reducing the number of beam-splitters, acceptable success probability and fidelity can be achieved. Therefore, the use of a three-dimensional quantum channel is advantageous, as it can maintain fidelity and success probability with a lower number of beam-splitters, even in the presence of noise.\\
	
	Finally, in Fig.~\ref{fig:all noise}, for $N = 3$ and three different values of $\alpha = 0.2, 0.5, 1$, the teleportation fidelity and success probability are investigated as functions of the noise probability $P_{\text{noise}}$ for all three types of quantum noise discussed. As can be seen, phase-flip noise has the least destructive effect in all cases. This is because, for the range of input coherent state amplitudes considered, the dominant term in the input state of the teleporters is the vacuum state $\left| 0 \right\rangle$, which, under phase-flip noise, remains unchanged as shown in Appendix~\ref{A}. Furthermore, as previously explained regarding the effect of bit-flip noise at small $\alpha$, this type of noise produces the largest reduction in both fidelity and success probability for small input amplitudes.

	\label{section4}
	\section{Conclusion}
	In this paper, we propose a scheme that, as a continuation of~\cite{8}, leads to an improvement in fidelity in the continuous-variable teleportation process while maintaining an acceptable success probability. In this scheme, by increasing the Hilbert space dimension of each teleporter, high fidelity can be achieved with a reduced number of teleporter arms, while the teleportation success probability is preserved or even enhanced. In the ideal noise-free case, it was found that for all input states, increasing the channel dimension, from two to three, improves both key performance metrics of the teleportation process, namely, the success probability and fidelity.\\
	
	Furthermore, the performance of the proposed protocol was investigated under realistic conditions, considering the effects of bit-flip, phase-flip, and depolarizing quantum noises. Overall, it was observed that the presence of noise degrades the teleportation performance, and for small input amplitudes, phase-flip noise causes the lowest disruption. It was also observed that, for small input amplitudes, increasing the number of beam-splitters reduces the teleportation success probability for all three types of the noises. Therefore, by employing a smaller number of beam-splitters, the teleportation process can be implemented under realistic noisy conditions with acceptable fidelity and success probability.
	
	\section*{Declaration of competing interest}
	The authors declare that they have no known competing financial interests or personal relationships that could have appeared to
	influence the work reported in this paper.
	
	\appendix
	\section{Noise Modelling}
	\label{A}
	In this section, Kraus operators for three common noises associated to qtrit, phase-flip, bit-flip and depolarizing noises are presented~\cite{13}.\\
	
	\subsection{Bit-Flip Noise}
	In quantum information processing, bit-flip noise refers to noise that causes one state to be converted to another. In our case, this noise acts on a qutrit state and converts it to another qutrit, with a finite probability. Mathematically, it can be imagined as follows:
	\begin{equation}
		\begin{array}{l}
			\left| 0 \right\rangle  \to \left| 1 \right\rangle ,\left| 1 \right\rangle  \to \left| 2 \right\rangle ,\left| 2 \right\rangle  \to \left| 0 \right\rangle\,, \\
			\left| 0 \right\rangle  \to \left| 2 \right\rangle ,\left| 1 \right\rangle  \to \left| 0 \right\rangle ,\left| 2 \right\rangle  \to \left| 1 \right\rangle \,,
		\end{array}
	\end{equation}
	The matrix representation of Kraus operators for bit-flip noise are
	\begin{equation}
		\begin{aligned}
			K_0 &= \sqrt{1 - P_{\rm noise}}\, \mathbb{1}\,, \\[1mm]
			K_1 &= \sqrt{\frac{P_{\rm noise}}{2}}\!
			\begin{pmatrix}
				0 & 0 & 1\\
				1 & 0 & 0\\
				0 & 1 & 0
			\end{pmatrix} \,,\\[1mm]
			K_2 &= \sqrt{\frac{P_{\rm noise}}{2}}\!
			\begin{pmatrix}
				0 & 1 & 0\\
				0 & 0 & 1\\
				1 & 0 & 0
			\end{pmatrix}\,,
		\end{aligned}
	\end{equation}\\
	where $\mathbb{1}=\begin{pmatrix}
		1 & 0 & 0\\
		0 & 1 & 0\\
		0 & 0 & 1
	\end{pmatrix}$ is the identity matrix.
	
	\subsection{Phase Flip Noise}
	The phase-flip noise affects the phase of the qutrit, in such a way that it transforms the qtrit state with a certain probability as follows:
	\begin{equation}
		\left| 0 \right\rangle  \to \left| 0 \right\rangle ,\left| 1 \right\rangle  \to  - \left| 1 \right\rangle ,\left| 2 \right\rangle  \to  - \left| 2 \right\rangle \,,
	\end{equation} 
	The matrix representation of Kraus operators corresponding to this noise can be written as:
	\begin{equation}
		\begin{aligned}
			K_0 &= \sqrt{1 - P_{\rm noise}}\, \mathbb{1}\,, \\[1mm]
			K_1 &= \sqrt{\frac{P_{\rm noise}}{2}}
			\begin{pmatrix}
				1 & 0 & 0\\
				0 & -1 & 0\\
				0 & 0 & 1
			\end{pmatrix} \,,\\[1mm]
			K_2 &= \sqrt{\frac{P_{\rm noise}}{2}}
			\begin{pmatrix}
				1 & 0 & 0\\
				0 & 1 & 0\\
				0 & 0 & -1
			\end{pmatrix}\,.
		\end{aligned}
	\end{equation}\\
	
	\subsection{Depolarizing Noise}
	Depolarizing noise causes the quantum state of a qutrit to transform into a completely mixed state with a certain probability. The depolarizing noise Kraus operators for qutrits are as follows:
	
	\begin{equation}
		\begin{array}{l l}
			K_0 = \sqrt{1 - P_{\rm noise}} \mathbb{1} &\,,\, K_1 = \sqrt{\frac{P_{\rm noise}}{8}} D_1 \,,\\[1mm]
			K_2 = \sqrt{\frac{P_{\rm noise}}{8}} D_2 &\,,\, K_3 = \sqrt{\frac{P_{\rm noise}}{8}} D_1^2\,, \\[1mm]
			K_4 = \sqrt{\frac{P_{\rm noise}}{8}} D_1 D_2 &\,,\, K_5 = \sqrt{\frac{P_{\rm noise}}{8}} D_1^2 D_2\,, \\[1mm]
			K_6 = \sqrt{\frac{P_{\rm noise}}{8}} D_1 D_2^2 &\,,\, K_7 = \sqrt{\frac{P_{\rm noise}}{8}} D_1^2 D_2^2 \,,\\[1mm]
			K_8 = \sqrt{\frac{P_{\rm noise}}{8}} D_2^2 \,,&
		\end{array}
	\end{equation}
	where 
	\begin{equation}
		\begin{aligned}
			D_1 &= \begin{pmatrix}
				0 & 1 & 0\\
				0 & 0 & 1\\
				1 & 0 & 0
			\end{pmatrix}, \\[4pt]
			D_2 &= \begin{pmatrix}
				1 & 0 & 0\\
				0 & \omega & 0\\
				0 & 0 & \omega^2
			\end{pmatrix}\,,
		\end{aligned}
	\end{equation}
	
	with $\omega  = {e^{i\frac{{2\pi }}{3}}}$.\\
	\section*{Data availability}
	Data will be made available on request.
	
	\clearpage

    \bibliographystyle{ieeetr}  %

	
\end{document}